\documentclass[twocolumn, resetfootnote]{aastex702}

\newcommand{\sersic}{S\'ersic }
\newcommand{\pysersic}{\texttt{pysersic} }
\newcommand{\statmorph}{\texttt{statmorph} }
\newcommand{\bagpipes}{\texttt{BAGPIPES} }
\newcommand{\pixedfit}{\texttt{piXedfit} }
\shorttitle{MQGs at Cosmic Noon: Morphology \& SF} 
\shortauthors{Prasal et al.}

\begin{document}

\title{Morphological and Star Formation Properties of Cosmic Noon Massive Quiescent Galaxies}

\author[orcid=0009-0007-4725-1455,sname=Prasal,gname=Vaidik]{Vaidik Prasal}
\affiliation{Indian Institute of Science Education and Research Pune, Dr. Homi Bhabha Road, Pashan, Pune 411008, India}
\email{vaidik.prasal@students.iiserpune.ac.in}

\author[orcid=0000-0002-1345-7371,sname=Wadadekar,gname=Yogesh]{Yogesh Wadadekar}
\affiliation{National Centre for Radio Astrophysics – Tata Institute of Fundamental Research, Post Bag 3, Ganeshkhind, Pune 411007, India}
\email{yogesh@ncra.tifr.res.in}

\author[orcid=0009-0002-2173-0953,sname=Biswas,gname=Pralay]{Pralay Biswas}
\affiliation{National Centre for Radio Astrophysics – Tata Institute of Fundamental Research, Post Bag 3, Ganeshkhind, Pune 411007, India}
\email{pbiswas@ncra.tifr.res.in}

\author[orcid=0009-0002-9692-3899,sname=Jain,gname=Rashi]{Rashi Jain}
\affiliation{National Centre for Radio Astrophysics – Tata Institute of Fundamental Research, Post Bag 3, Ganeshkhind, Pune 411007, India}
\email{rjain@ncra.tifr.res.in}

\email[show]{vaidik.prasal@students.iiserpune.ac.in}




\begin{abstract}
We analyze the star formation and morphological properties of massive quiescent galaxies at cosmic noon ($2 < z < 3$) in the Abell 2744 field, using deep JWST NIRCam \added[changed from broad-band]{broadband} and medium-band imaging from the UNCOVER Treasury program and the MegaScience survey, complemented by archival HST data. Using \bagpipes SED modeling, we select 14 unique massive quiescent galaxies ($M_* \gtrsim 10^{10}$ M$_\odot$, $\mathrm{sSFR} < 0.2/t_\mathrm{age}$). Morphological analysis with \statmorph and \pysersic reveals that most galaxies are intermediate type or S0s with a median \sersic index $n \sim 4$, consistent with bulge-dominated systems. This value remains constant over $z \sim 1.5$--$4$, indicating that the morphology of massive galaxies is linked to their quiescence since at least $z \sim 4$. Spatially resolved SED modeling with \pixedfit shows that \added[percentage removed]{11 out of 14} galaxies exhibit positive radial sSFR gradients, providing direct evidence for inside-out quenching, with the mean sSFR increasing by $\sim2$ dex from $R/R_e = 0.5$ to $4.5$. Formation time ($t_{50}$) profiles confirm that inner regions formed $\approx 0.5$ Gyr earlier, on average, than the outer regions, and quenching timescale profiles show that the cores were quenched more rapidly than the outskirts. Most galaxies are compact, with a mean half-mass radius of $R_e = 1.95 \pm 0.13$ kpc. \added[Ref. comment: soften AGN claims in abstract]{The observed inside-out quenching pattern is consistent with quenching driven by AGN feedback, while the bulge-dominated morphologies suggest morphological quenching may also contribute.}
\end{abstract}



\section{Introduction}
Massive quiescent galaxies at high redshifts ($z > 2$) challenge our understanding of galaxy evolution and serve as critical laboratories for testing galaxy formation and evolution models \citep{Cimatti2004, Glazebrook2004}. These systems are unusual because they have built up a large stellar mass and have thereafter stopped forming stars by the time the Universe was about 3 billion years old, a period when star formation was generally highly active \citep{Madau2014}. Understanding how, when, and why these galaxies quenched remains one of the central open questions in extragalactic astronomy. The existence of massive quiescent galaxies at high redshifts poses a problem for many theoretical models of galaxy formation because they require extremely rapid formation and quenching mechanisms to assemble large stellar masses (often $> 10^{11}$ M$_\odot$) and then to completely reverse the star-formation process, all within the first few billion years after the Big Bang. 

Cosmological hydrodynamical simulations such as IllustrisTNG \citep{Pillepich2018}, EAGLE \citep{Schaye2015}, SIMBA \citep{Dave2019}, and FLAMINGO \citep{Schaye2023} make predictions about the number densities of massive quiescent galaxies that can be compared with observations. Recent observations with the James Webb Space Telescope (JWST) have found number densities of massive quiescent galaxies at $z > 3$ that exceed predictions from even the most recent semi-analytic and hydrodynamic models \citep{Carnall2023a,Valentino2023,Baker2025}. This tension between simulations and observations suggests that models may underestimate the efficiency of early mass assembly and quenching processes, or that the prescriptions for star formation and feedback at early times require revision \citep[but see also \citealt{GuzmanOrtega2025} for reconciliations using dust-aware mock observations]{deGraaff2025}.

To address these tensions and explain the rapid shutdown of star formation, several physical mechanisms have been proposed. Active galactic nucleus (AGN) feedback is widely considered the leading candidate for quenching massive galaxies, operating in both ``quasar mode'' (radiatively driven winds) and ``jet/radio mode'' (mechanical energy injection) \citep{Croton2006,Bower2006,Fabian2012}. Recent JWST observations of mass outflows from cosmic noon massive quiescent galaxies provide direct evidence for AGN-driven feedback as a quenching mechanism \citep{Park2024,Belli2024,Scholtz2026}. A complementary mechanism is morphological quenching, where the growth of a massive stellar bulge stabilizes the gas disk against fragmentation and collapse, effectively shutting down star formation even when cold gas is present \citep{Martig2009}.

The physical mechanisms driving this rapid quenching are closely tied to the structural evolution of these galaxies. A defining feature of massive quiescent galaxies at $z>2$ (extending to $z\sim 4$) is their extreme compactness, which is most accurately revealed in the rest-frame near-infrared where stellar mass distributions, dominated by low mass stars, are more clearly traced \citep{Martorano2024,Haryana_2025}. These systems exhibit typical effective radii of $R_e\sim$ 0.6--1.5 kpc, a factor of 4 to 6 smaller than local ellipticals of comparable mass, resulting in stellar mass surface densities approximately two orders of magnitude higher than their low-redshift counterparts \citep{vanDokkum2008,Martorano2024,Haryana_2025}. 

A recent spatially resolved analysis \citep{Haryana_2025} indicates that while their central 1 kpc cores were already as dense as local quiescent galaxies by $z\sim 4$, their subsequent size evolution is driven by a two-phase assembly process: initial growth dominated by major mergers at $z \gtrsim 2$, followed by minor dry mergers at $z \lesssim 2$ that primarily build up the galaxy outskirts while leaving the dense core intact. Cosmological simulations also support this size evolution driven primarily by minor mergers \citep{Naab2009,Oser2012,Wellons2015,Rodriguez-Gomez2016}. 

Consistent with their compact sizes and centrally concentrated mass distributions, massive quiescent galaxies at $z > 2$ typically exhibit high \sersic indices ($n \sim 3$--6), indicating concentrated, bulge-dominated light profiles similar to local elliptical galaxies \citep{vanDokkum2010,vanderWel2014,Cutler2024}. This suggests that the central bulge component was already in place by $z \sim 2$--3, with galaxies subsequently growing their envelopes through dry merging.

An alternate way for quantifying galaxy morphologies, particularly for high-redshift galaxies with small angular size, is by using non-parametric measures such as the CAS (Concentration--Asymmetry--Smoothness) system \citep{Abraham1996,Conselice2003} and the Gini--$M_{20}$ plane \citep{Lotz2004,Lotz2008}. Quiescent galaxies at $z > 2$ tend to occupy the high-concentration, low-asymmetry region of the CAS parameter space \citep{Conselice2003,Lee2013,Yao2023}, consistent with their smooth, symmetric morphologies \citep{Bershady2000,Lee2013,Kartaltepe2023}. In the Gini--$M_{20}$ plane, they fall in the ``bulge-dominated (E/S0)'' region \citep{Lotz2008,Lee2013}. While these global morphological measures reveal the overall structural properties of quiescent galaxies, understanding the spatial distribution of their stellar populations requires spatially resolved analyses.

Spatially resolved studies of massive quiescent galaxies provide critical constraints on how star formation shuts down within individual systems. The paradigm of inside-out quenching, where star formation is first suppressed in the central regions while the outskirts remain active, has emerged from theoretical predictions and observational evidence spanning from the local universe to high redshifts. 

Observations of massive galaxies at $z \sim 2.2$ reveal the inside-out quenching process, where star formation ceases in the dense inner regions on timescales of $\lesssim 1$ Gyr while outer disks remain active, rapidly building central mass densities comparable to local early-type galaxies \citep{Tacchella2015}. This is supported by high-resolution molecular gas imaging at $z \sim 2$, which shows centrally suppressed gas fractions and short depletion times in the inner $1$--$2$ kpc \citep{Spilker2019}. Similar trends persist in the local universe; integral field surveys like the Calar Alto Legacy Integral Field Area   (CALIFA) survey and the Mapping Nearby Galaxies at Apache Point Observatory (MaNGA) survey confirm that massive galaxies preferentially exhibit outwardly increasing sSFR profiles, with central bulges quenching significantly faster than their disks \citep{Delgado2016,Lin2019}. Furthermore, \citet{Lin2019} found a strong link between dense, quenched cores and bulge-dominated structures (high \sersic indices) across all environments, demonstrating that central bulge growth plays a key role in shutting down star formation. Theoretically too, cosmological simulations such as IllustrisTNG predict this inside-out quenching pattern as a natural consequence of low-accretion kinetic AGN feedback. In this framework, central kinetic winds evacuate gas from the galactic core, creating positive sSFR radial gradients that are quantitatively consistent with observations up to $z \sim 1$ \citep{Nelson2019, Nelson2021}. Taken together, these studies suggest that a combination of central bulge growth and AGN-driven winds drives the rapid central quenching of massive galaxies across cosmic time.

With JWST, spatially resolved spectral energy distribution (SED) modeling \citep[see][for a review on SED modeling]{Conroy2013,Iyer2025} has become possible at $z > 2$ due to JWST's unprecedented sensitivity and resolution in the near-infrared, enabling pixel-level stellar population analysis. At this redshift, the near-infrared observations have an additional advantage;  they give us access to the rest-frame optical wavelengths. Recent JWST studies have demonstrated that the majority of massive quiescent galaxies at $z \sim 2$--3 show positive sSFR gradients, lower sSFR in the center, and higher sSFR in the outskirts, directly confirming inside-out quenching \citep{Haryana_2025,Laishram2025}.

In this paper, we present a comprehensive analysis of the morphological and star formation properties of massive quiescent galaxies ($M_* \gtrsim 10^{10}$~M$_\odot$) at cosmic noon ($2 < z < 3$), leveraging deep JWST NIRCam imaging from the Ultradeep NIRSpec and NIRCam ObserVations before the Epoch of Reionization (UNCOVER) Treasury program \citep{Bezanson2024} and the ``Medium Bands, Mega Science" \citep[MegaScience]{Suess2024} medium-band survey of the Abell 2744 lensing cluster field. We select quiescent galaxies using the Bayesian Analysis of Galaxies for Physical Inference and Parameter EStimation \citep[\texttt{BAGPIPES}]{Carnall2018} code for the SED modeling of the combined HST + JWST photometry with an sSFR criterion, perform morphological analysis using \statmorph \citep[non-parametric CAS, Gini, $M_{20}$]{Rodriguez-Gomez2019} and \pysersic \citep[two-component bulge and disk decomposition]{Pasha2023} across all NIRCam broadband filters, and probe the spatial distribution of stellar populations through spatially resolved SED modeling with \texttt{piXedfit} \citep{Abdurrouf2021}. We derive pixel-level maps and radial profiles of stellar mass, star formation rate (SFR), sSFR, formation time ($t_{50}$), quenching timescale ($t_q - t_{50}$), and the rest-frame U$-$V and V$-$J colors to constrain the quenching history.

A key strength of our SED modeling lies in the combination of broadband and medium-band photometry. The inclusion of medium bands provides tighter constraints on photometric redshifts \citep{Suess2024} and mitigates the systematic biases in stellar mass and SFR estimates that frequently affect broadband-only analyses \citep{Martis2025}.

This paper is organized as follows. Section~\ref{sec:dataset} describes the UNCOVER and MegaScience datasets. Section~\ref{sec:methods} details our sample selection, morphological analysis, and spatially resolved SED modeling methodology. Section~\ref{sec:results} presents the morphological properties, \pixedfit results, and radial profiles. Section~\ref{sec:discussion} discusses the implications of our findings in the context of inside-out quenching and galaxy evolution at cosmic noon. Section~\ref{sec:conclusions} summarizes our main conclusions. We have used the \texttt{Planck2018} cosmology \citep[$H_0 = 67.66$ km s$^{-1}$ Mpc$^{-1},\, \Omega_m = 0.31,\, \Omega_{\Lambda} = 0.69$]{Planck2018} throughout this work. 

\begin{deluxetable*}{lcl}[ht!]
\tabletypesize{\small}
\tablecaption{BAGPIPES fitting parameters\label{tab:bagpipesparams}}
\tablehead{
\colhead{Parameter} & \colhead{Value / Range} & \colhead{Description}
}
\setlength{\tabcolsep}{20pt}
\startdata
\cutinhead{Star Formation History (Delayed Tau Model)}
age & [0.1, 15] & Time since star formation began in Gyr \\
tau & [0.3, 10] & e-folding timescale of SFH decay in Gyr \\
massformed & [4, 15]  & Log of total stellar mass formed wrt solar mass \\
metallicity & [0.1, 2.5] & Log of stellar metallicity wrt solar metallicity\\
\cutinhead{Dust Attenuation}
type & ``Calzetti" & Dust attenuation law \\
eta & 2 & Multiplicative factor for the Calzetti law \\
$A_V$ & [0, 6] & V-band attenuation in mag \\
\cutinhead{Nebular Emission}
logU & $-3$ & Log of the ionization parameter \\
\cutinhead{Fit Instructions}
redshift & [0, 15] & Redshift range for the fit \\
\enddata
\end{deluxetable*}

\section{Dataset} \label{sec:dataset}

We utilize image mosaics, weight maps, and photometric products from the coordinated JWST surveys of the massive lensing cluster Abell 2744 ($z=0.308$). This dataset is primarily composed of observations from the UNCOVER JWST Cycle 1 Treasury program \citep[PIs Labb\'e \& Bezanson; GO-2561;][]{Bezanson2024} and the MegaScience JWST Cycle 2 program \citep[PI Suess; GO-4111;][]{Suess2024}.

\subsection{UNCOVER Survey}

The primary UNCOVER NIRCam mosaic was imaged through six broadband filters (F115W, F150W, F200W, F277W, F356W, and F444W) and one medium-band filter (F410M), achieving imaging depths of $\sim29$--30 AB \citep{Bezanson2024}. UNCOVER also obtained ultradeep NIRSpec/PRISM spectroscopy ($R\sim30$--300) spanning $0.6$--$5.3\ \mu\mathrm{m}$ for approximately 500 unique galaxies. These spectroscopic observations utilized a multi-mask strategy, with total integration times ranging from 2.7 to 17.4 hours, to reach continuum depths of $\sim29$ AB \citep{Price2025}.

\subsection{Medium Bands, Mega Science Survey}

To complement the broadband imaging, we incorporate data from the MegaScience survey, which provides comprehensive medium-band coverage of the same field \citep{Suess2024}. MegaScience delivered $29.2\ \mathrm{arcmin}^2$ of deep NIRCam imaging (up to $\sim30$ AB) through 11 additional medium-band filters (F140M, F162M, F182M, F210M, F250M, F300M, F335M, F360M, F430M, F460M, and F480M) and two short-wavelength broadband filters (F070W and F090W). When combined with UNCOVER, this results in a unique dataset featuring complete coverage across all 20 NIRCam broadband and medium-band filters. The inclusion of medium-band photometry reduces photometric redshift scatter ($\sigma_{\mathrm{NMAD}}$) and catastrophic outlier rates by factors of 2--3 compared to broadband-only measurements \citep{Suess2024}.

\subsection{UNCOVER SUPER Catalog (DR3)}

We utilize the value-added UNCOVER SUPER Catalog from the Data Release 3\footnote{\url{https://jwst-uncover.github.io/DR3.html}} (DR3), which provides multi-wavelength photometry for 74,020 sources in the Abell 2744 field \citep{Weaver2024, Suess2024}. The UNCOVER team constructed this catalog using a $56.2\ \mathrm{arcmin}^2$ detection image, created from a noise-weighted stack of long-wavelength filters (F277W, F356W, and F444W). The co-added mosaics incorporate data from several other public JWST programs in the 
Abell 2744 field, including GLASS-ERS \citep{Treu2022}, DDT-2756 (PI Chen), ALT \citep{Naidu2024}, MAGNIF (PI Sun), and GO-3538 (PI Iani). The catalog also integrates archival Hubble Space Telescope imaging from the Hubble Frontier Fields (HFF) and BUFFALO programs, spanning seven filters (F435W, F606W, F814W, F105W, F125W, F140W, and F160W).

To ensure accurate SED modeling, the UNCOVER team matched all images to the F444W point-spread function (PSF) using convolution kernels. Finally, the strong lensing magnifications for all galaxies in the catalog are based on the updated mass models of \citet{Furtak2023} and \citet{Price2025} (taken from the DR4\footnote{\url{https://jwst-uncover.github.io/DR4.html}}), which utilize the deep JWST imaging to provide significantly improved constraints.

\subsection{DAWN JWST Archive}

The DAWN JWST Archive\footnote{\url{https://dawn-cph.github.io/dja/index.html}} (DJA) hosts publicly released JWST galaxy data reduced with the \texttt{grizli} \citep{Grizli2019, Grizli2021} and \texttt{msaexp} \citep{brammer_msaexp} reduction pipelines. We have accessed the catalog of the galaxies with grade 3 NIRSpec data from the DJA \citep{brammer_msaexp,deGraaff2024,Heintz2024}. A grade 3 spectrum implies a robust redshift from one or more emission or absorption features. Out of the 37,528 grade 3 spectra available (as on 29th December 2025), 5 are common with our sample. For these objects, we use the spectroscopic redshifts for SED modeling, as elaborated in the next section. 

\section{Methods} \label{sec:methods}
\subsection{Stellar Population Synthesis with BAGPIPES}

To determine the global physical properties of galaxies, such as stellar mass, star formation rate, and photometric redshift, we use SED modeling. By comparing the observed photometry in various filters to synthetic spectra produced by stellar population synthesis models, we can derive joint posterior distributions for these critical parameters, which, in turn, form the basis of our quiescent galaxy sample selection.

We perform SED modeling using data from the JWST NIRCam filters (broadband and medium bands), HST/WFC3 filters (F105W, F125W, F140W, F160W), and HST/ACS filters (F435W, F606W, F814W), where available, using the \texttt{BAGPIPES} framework \citep{Carnall2018}. \bagpipes provides a flexible framework for generating synthetic galaxy spectra from user-specified parameters and for fitting those models to observational data. The framework incorporates BC03 SPS models from the 2016 update of \citet{Bruzual2003}, combined with the \added[citation corrected]{\citet{Kroupa2001}} initial mass function. Users specify prior distributions and permissible ranges for physical parameters, as well as parametric forms for star formation histories. \bagpipes employs the PyMultiNest nested sampling algorithm \citep{Feroz2008, Feroz2009, Feroz2019, Buchner2014} to efficiently sample the parameter space and generate model SEDs. By extensively sampling the posterior distribution, \bagpipes simultaneously constrains the photometric redshift and SPS parameters, and returns joint posterior estimates of redshift, stellar mass, star formation rate, gas-phase metallicity, and dust attenuation. \added{We use the delayed tau Star Formation History (SFH) and the Nov 2022 JWST filter curves from \texttt{sedpy} \citep{johnson_2021_4582723} for our analysis.} For each galaxy, we mask the filters with $\textrm{Signal-to-Noise Ratio (SNR)} \leqslant 3$. We also divide the fluxes by their strong lensing magnifications ($\mu$, taken from the catalog) to correct for strong lensing. For the galaxies having a spectroscopic redshift (from DJA), we fix their redshifts to the spectroscopic value and then run \bagpipes on them. See Table~\ref{tab:bagpipesparams} for the parameters and priors used for the \bagpipes modeling. 

\subsection{Sample selection using BAGPIPES results}

Having derived the physical properties of the galaxies in the UNCOVER catalog using \texttt{BAGPIPES}, we then identify massive galaxies at cosmic noon that have already ceased significant star formation from the output catalog as follows:

\begin{itemize}
    \item Stellar Mass: We select galaxies with stellar masses (50th percentile) $M_* \gtrsim 10^{10}$ M$_\odot$ to focus on the most massive systems.
    \item sSFR: We classify galaxies with
    \begin{equation}
        \mathrm{sSFR} < \frac{0.2}{t_{\mathrm{age}}},
        \label{eq:ssfr_cut}
    \end{equation}
    where $t_{\mathrm{age}}$ is the age of the universe at the galaxy's redshift, as quiescent, indicating minimal ongoing star formation \citep{Franx2008, Gallazzi2014, Schreiber2018, Carnall2023a, Baker2025}. 
    \item Redshift Range: We restrict our sample to galaxies within the redshift range $2 < z < 3$ to target the epoch of interest for cosmic noon studies.
\end{itemize}

\begin{figure}[h!]
\centering
\includegraphics[width=\columnwidth]{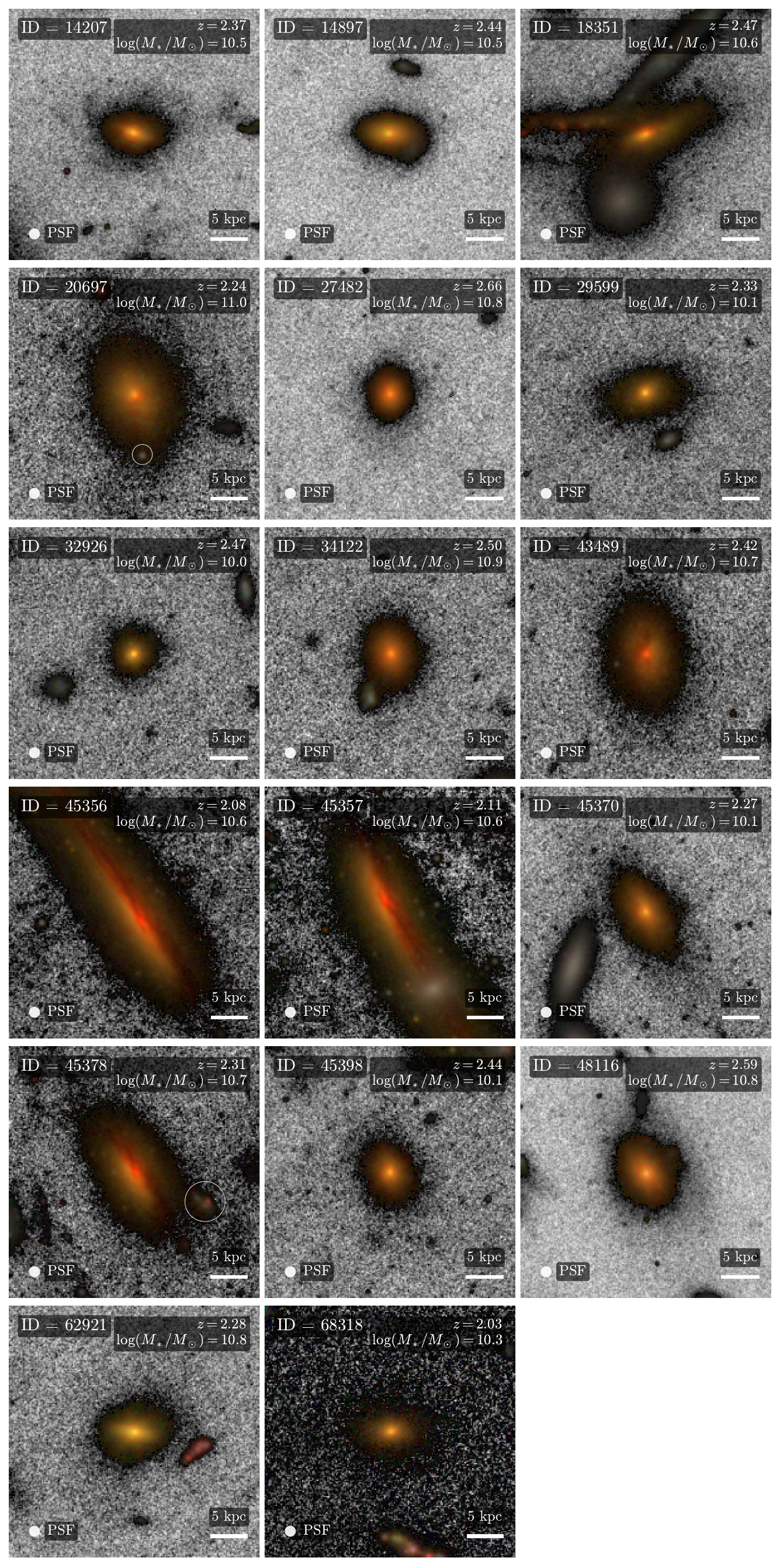}
\caption{Our sample of 17 massive quiescent galaxies is shown here as $4''\times 4''$ RGB images with filters F200W, F150W, and F115W, centered on the galaxies. Three of these galaxies (\texttt{ID\_DR3} = 45356, 45357, 45378) are multiply-lensed images of the same source, and two of them (\texttt{ID\_DR3} = 45370, 45398) are multiply-lensed images of another source \citep{Furtak2023}. The white circle at the bottom left represents the PSF size of the F444W filter image (FWHM = 0\farcs176). This is $\approx21\%$ higher than the empirical PSF \citep{JDox2016}. UNCOVER uses pixel scales of 0.02\arcsec\ and 0.04\arcsec\ for short wavelength (SW) and long wavelength (LW) filters, respectively. We have also circled the foreground galaxies near galaxy \texttt{ID\_DR3} = 20697 and near galaxy \texttt{ID\_DR3} = 45378. See the text for details. }  
\label{fig:rgb_collage}
\end{figure}

\begin{figure}
\centering
\includegraphics[width=0.8\columnwidth]{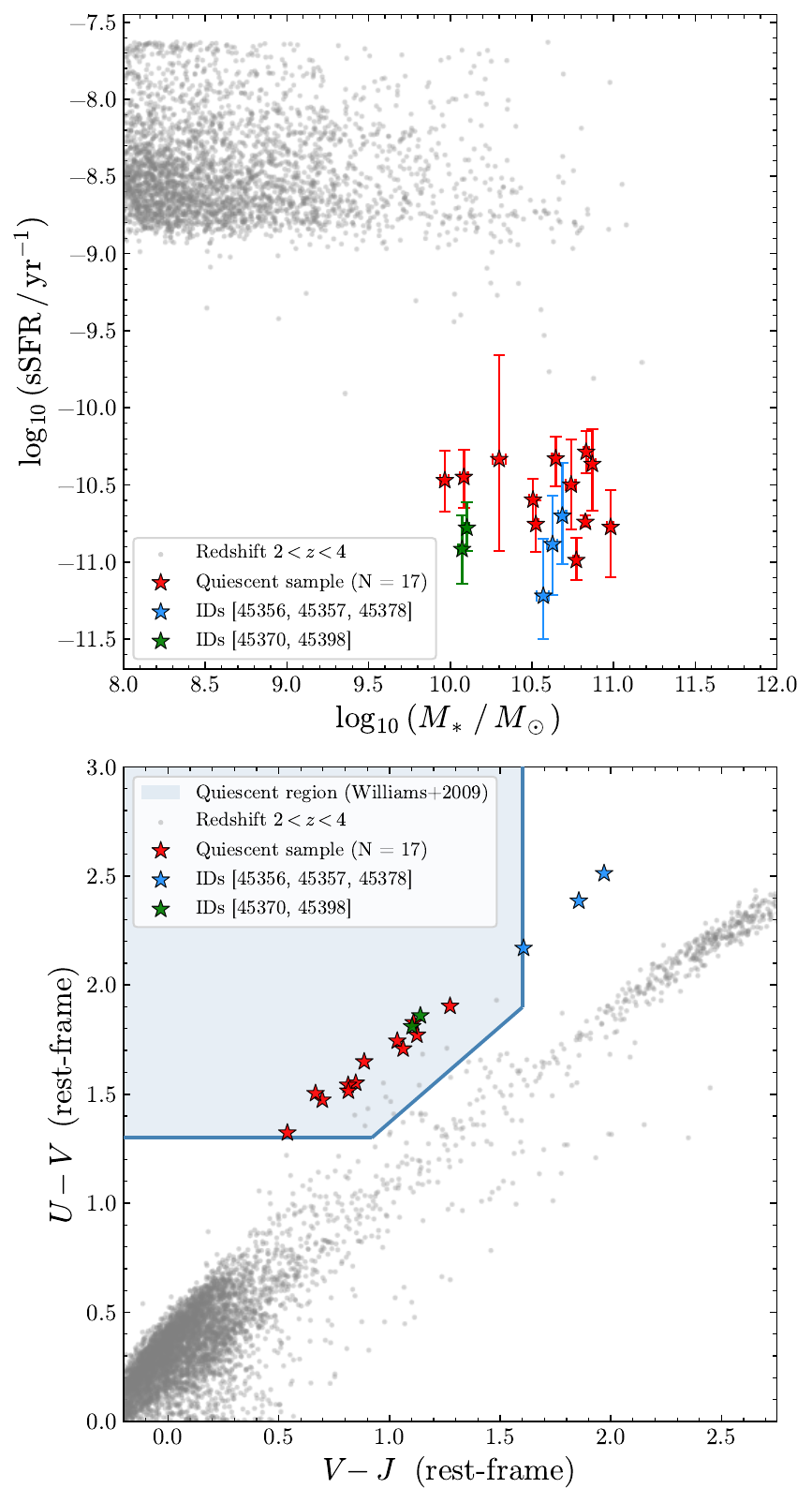}
\caption{The stellar mass vs sSFR (in log scale, top panel) and $UVJ$ diagram (bottom panel) for our sample of 17 massive quiescent galaxies out of all the galaxies in the survey with redshifts between 2 and 3. The quiescent selection boundary from \citet{Williams2009} is shown in the $UVJ$ plot, as a light-blue shaded region. The stellar mass vs sSFR plot shows that our sample galaxies lie well below the star-forming galaxies, consistent with their quiescent nature. The three multiply-lensed images of the first source (\texttt{ID\_DR3} = 45356, 45357, 45378) are marked with blue stars, and the two multiply-lensed images of the second source (\texttt{ID\_DR3} = 45370, 45398) are marked with green stars. The gray points represent high-redshift galaxies in the redshift range $2<z<4$ in the UNCOVER survey.}
\label{fig:sfms_uvj}
\end{figure}

After applying these selection criteria, we identify a sample of 17 massive quiescent galaxies for further analysis, shown in Figure~\ref{fig:rgb_collage}. Figure~\ref{fig:sfms_uvj} shows the stellar mass vs sSFR plot and UVJ diagram for our sample. Among these, \citet{Furtak2023} identify 3 (\texttt{ID\_DR3} = 45356, 45357, 45378) as the multiply-imaged source with ID 67, and 2 (\texttt{ID\_DR3} = 45370, 45398) as the multiply-imaged source with ID 69. So effectively, we have 14 unique massive quiescent galaxies in our sample. We use the least magnified image (\texttt{ID\_DR3} = 45378) of the lensed source ID 67 in our analysis, since the lensing has highly sheared and distorted the other two images \citep{Siegel2025}. Similarly, we use the less magnified image (\texttt{ID\_DR3} = 45398) of the lensed source ID 69 in our analysis. We note that the galaxies with the source ID 67 lie just outside or a little bit outside of the quiescent region defined by \citet{Williams2009}, see Figure~\ref{fig:sfms_uvj}. Nevertheless, we classify them as quiescent due to their extremely low sSFR, as shown in the mass--sSFR plot. Table~\ref{tab:final_samples} presents the estimated SPS parameters of the final sample.

\added{In the literature, there are two popular choices for the choice of Star Formation Histories (SFHs): the delayed tau model \citep{Sato2024, Haryana_2025}, and the double power law model \citep{Carnall2018, Carnall2024, Baker2025}. We choose the delayed tau model for our analysis. We compared the derived parameters for the galaxies in our sample with those derived using the double power-law SFH, and found no significant differences. To assess the robustness of our methodology, we also test a non-parametric SFH following \citet{Leja2019}. \bagpipes does not support non-parametric SFH for galaxies without a spectroscopic redshift. So, we test the 5 galaxies with spectroscopic redshifts (\texttt{ID\_DR3} = 27482, 34122, 45378, 45398, and 48116) using the non-parametric SFH and find that all of them satisfy the quiescence criterion (See Equation~\ref{eq:ssfr_cut}).}

\begin{deluxetable*}{cccccccccccc}[ht!]
\tabletypesize{\scriptsize}
\tablecaption{Physical properties of final sample galaxies from BAGPIPES SED fitting\label{tab:final_samples}}
\tablehead{
\colhead{ID} & \colhead{Age} & \colhead{log($M_{\mathrm{formed}}/M_\odot$)} & \colhead{Metallicity} & \colhead{$\tau$} & \colhead{$A_V$} & \colhead{Redshift} & \colhead{log($M_*/M_\odot$)} & \colhead{SFR} & \colhead{log(sSFR/yr$^{-1}$)} & \colhead{MW age} & \colhead{$\mu$} \\
\colhead{} & \colhead{(Gyr)} & \colhead{} & \colhead{} & \colhead{(Gyr)} & \colhead{(mag)} & \colhead{} & \colhead{} & \colhead{($M_\odot$/yr)} & \colhead{} & \colhead{(Gyr)} & \colhead{}
}
\startdata
14207 & $2.55_{-0.17}^{+0.14}$ & $10.79_{-0.02}^{+0.02}$ & $0.258_{-0.074}^{+0.164}$ & $0.32_{-0.01}^{+0.03}$ & $0.47_{-0.13}^{+0.12}$ & $2.37_{-0.06}^{+0.08}$ & $10.52_{-0.02}^{+0.02}$ & $0.59_{-0.20}^{+0.25}$ & $-10.76_{-0.18}^{+0.15}$ & $1.92_{-0.16}^{+0.12}$ & $2.09_{-0.03}^{+0.00}$ \\
14897 & $2.37_{-0.14}^{+0.15}$ & $10.77_{-0.03}^{+0.03}$ & $0.224_{-0.071}^{+0.112}$ & $0.31_{-0.01}^{+0.02}$ & $0.29_{-0.12}^{+0.10}$ & $2.44_{-0.07}^{+0.05}$ & $10.51_{-0.03}^{+0.02}$ & $0.83_{-0.24}^{+0.30}$ & $-10.60_{-0.15}^{+0.14}$ & $1.75_{-0.13}^{+0.13}$ & $2.33_{-0.03}^{+0.01}$ \\
18351 & $2.23_{-0.20}^{+0.19}$ & $10.90_{-0.03}^{+0.02}$ & $0.258_{-0.067}^{+0.119}$ & $0.33_{-0.02}^{+0.03}$ & $0.52_{-0.13}^{+0.12}$ & $2.47_{-0.04}^{+0.04}$ & $10.65_{-0.02}^{+0.02}$ & $2.04_{-0.66}^{+0.82}$ & $-10.33_{-0.18}^{+0.14}$ & $1.59_{-0.16}^{+0.16}$ & $2.08_{-0.02}^{+0.00}$ \\
20697 & $2.65_{-0.29}^{+0.18}$ & $11.25_{-0.03}^{+0.02}$ & $0.391_{-0.181}^{+0.851}$ & $0.32_{-0.02}^{+0.03}$ & $0.70_{-0.45}^{+0.16}$ & $2.24_{-0.04}^{+0.07}$ & $10.98_{-0.02}^{+0.02}$ & $1.64_{-0.89}^{+1.13}$ & $-10.77_{-0.32}^{+0.24}$ & $1.98_{-0.26}^{+0.18}$ & $1.63_{-0.01}^{+0.00}$ \\
27482$^{*}$ & $2.39_{-0.03}^{+0.02}$ & $11.09_{-0.01}^{+0.01}$ & $0.194_{-0.024}^{+0.015}$ & $0.30_{-0.00}^{+0.00}$ & $0.78_{-0.04}^{+0.06}$ & 2.66 & $10.83_{-0.01}^{+0.01}$ & $1.23_{-0.09}^{+0.13}$ & $-10.74_{-0.03}^{+0.05}$ & $1.79_{-0.03}^{+0.02}$ & $1.80_{-0.02}^{+0.01}$ \\
29599 & $2.30_{-0.19}^{+0.25}$ & $10.34_{-0.03}^{+0.03}$ & $0.242_{-0.076}^{+0.161}$ & $0.32_{-0.02}^{+0.03}$ & $0.45_{-0.13}^{+0.13}$ & $2.33_{-0.05}^{+0.07}$ & $10.08_{-0.03}^{+0.02}$ & $0.43_{-0.16}^{+0.20}$ & $-10.45_{-0.20}^{+0.18}$ & $1.66_{-0.16}^{+0.19}$ & $2.23_{-0.05}^{+0.02}$ \\
32926 & $2.34_{-0.19}^{+0.20}$ & $10.23_{-0.03}^{+0.03}$ & $0.191_{-0.055}^{+0.076}$ & $0.33_{-0.02}^{+0.03}$ & $0.10_{-0.07}^{+0.13}$ & $2.47_{-0.07}^{+0.06}$ & $9.97_{-0.03}^{+0.02}$ & $0.30_{-0.10}^{+0.17}$ & $-10.47_{-0.20}^{+0.19}$ & $1.68_{-0.16}^{+0.18}$ & $1.73_{-0.04}^{+0.01}$ \\
34122$^{*}$ & $2.24_{-0.19}^{+0.22}$ & $11.12_{-0.03}^{+0.02}$ & $0.295_{-0.091}^{+0.387}$ & $0.32_{-0.02}^{+0.04}$ & $0.90_{-0.29}^{+0.14}$ & 2.50 & $10.87_{-0.02}^{+0.02}$ & $3.20_{-1.56}^{+2.05}$ & $-10.37_{-0.30}^{+0.23}$ & $1.60_{-0.18}^{+0.21}$ & $1.37_{-0.01}^{+0.00}$ \\
43489 & $2.43_{-0.29}^{+0.20}$ & $11.00_{-0.03}^{+0.03}$ & $0.435_{-0.193}^{+0.666}$ & $0.33_{-0.02}^{+0.04}$ & $0.98_{-0.29}^{+0.20}$ & $2.42_{-0.11}^{+0.11}$ & $10.74_{-0.03}^{+0.03}$ & $1.72_{-0.84}^{+1.71}$ & $-10.50_{-0.29}^{+0.29}$ & $1.76_{-0.26}^{+0.20}$ & $1.72_{-0.06}^{+0.03}$ \\
45378$^{*}$$^{\dagger}$ & $2.53_{-0.26}^{+0.20}$ & $10.95_{-0.03}^{+0.03}$ & $0.374_{-0.184}^{+1.400}$ & $0.32_{-0.01}^{+0.03}$ & $1.46_{-0.59}^{+0.25}$ & 2.31 & $10.69_{-0.02}^{+0.02}$ & $1.00_{-0.53}^{+1.14}$ & $-10.70_{-0.31}^{+0.34}$ & $1.88_{-0.25}^{+0.20}$ & $5.50_{-0.44}^{+0.34}$ \\
45398$^{*}$ & $2.54_{-0.14}^{+0.08}$ & $10.37_{-0.02}^{+0.02}$ & $0.242_{-0.053}^{+0.111}$ & $0.31_{-0.01}^{+0.02}$ & $0.81_{-0.13}^{+0.08}$ & 2.44 & $10.10_{-0.01}^{+0.02}$ & $0.21_{-0.06}^{+0.10}$ & $-10.78_{-0.15}^{+0.16}$ & $1.90_{-0.12}^{+0.08}$ & $6.23_{-0.64}^{+0.23}$ \\
48116$^{*}$ & $2.12_{-0.14}^{+0.18}$ & $11.09_{-0.02}^{+0.02}$ & $0.194_{-0.035}^{+0.046}$ & $0.32_{-0.02}^{+0.02}$ & $0.56_{-0.10}^{+0.11}$ & 2.59 & $10.83_{-0.02}^{+0.02}$ & $3.49_{-0.94}^{+1.27}$ & $-10.29_{-0.14}^{+0.13}$ & $1.50_{-0.12}^{+0.14}$ & $2.00_{-0.09}^{+0.07}$ \\
62921 & $2.68_{-0.12}^{+0.11}$ & $11.05_{-0.02}^{+0.02}$ & $0.205_{-0.030}^{+0.046}$ & $0.31_{-0.01}^{+0.02}$ & $0.16_{-0.08}^{+0.08}$ & $2.28_{-0.05}^{+0.04}$ & $10.77_{-0.02}^{+0.02}$ & $0.62_{-0.17}^{+0.22}$ & $-10.99_{-0.13}^{+0.15}$ & $2.05_{-0.11}^{+0.10}$ & $1.50_{-0.01}^{+0.00}$ \\
68318 & $2.57_{-0.58}^{+0.45}$ & $10.56_{-0.05}^{+0.05}$ & $0.373_{-0.185}^{+0.571}$ & $0.38_{-0.06}^{+0.14}$ & $0.75_{-0.44}^{+0.31}$ & $2.03_{-0.08}^{+0.18}$ & $10.30_{-0.04}^{+0.04}$ & $0.90_{-0.65}^{+3.25}$ & $-10.33_{-0.60}^{+0.67}$ & $1.78_{-0.57}^{+0.47}$ & $1.33_{-0.01}^{+0.00}$ \\
\enddata
\tablecomments{Values shown as median with 16th/84th percentile uncertainties: $\mathrm{value}_{-\sigma_{\mathrm{lo}}}^{+\sigma_{\mathrm{hi}}}$. All physical parameters have been derived from BAGPIPES SED fitting with delayed tau star formation history. $^{*}$ Indicates galaxies with confirmed spectroscopic redshifts, obtained from DJA; for these objects, the redshift error is not shown. $^{\dagger}$ The BAGPIPES fit is not reliable due to bimodal posteriors; see the Appendix~\ref{sec:lensed_galaxies} for more information.}
\end{deluxetable*}

\subsection{Morphological analysis using statmorph and pysersic}

With our final sample defined, we turn to characterizing their structural properties to understand their physical sizes, light distributions, and overall morphologies. To achieve a comprehensive view of their structures, we employ both non-parametric and parametric morphological modeling. We perform the non-parametric morphological analysis of the selected massive quiescent galaxies using the \statmorph package \citep{Rodriguez-Gomez2019}. \statmorph is a Python library designed for measuring different non-parametric (CASGM) morphological statistics of galaxies in astronomical images, including concentration, asymmetry, smoothness (CAS), Gini coefficient, and the M\textsubscript{20} parameter (GM). Taken together, these parameters provide insights into the structural properties of galaxies, such as their light distribution, symmetry, and clumpiness. For parametric morphological analysis, we use the \pysersic package \citep{Pasha2023} to perform bulge-disk decomposition of the galaxies. \pysersic is a Python library that fits \sersic (and other) profiles to astronomical images using Bayesian inference.  

Firstly, we prepare 12\arcsec\ cutouts of the broadband filter images (F070W, F090W, F115W, F150W, F200W, F277W, F356W, and F444W). We use the bCG-subtracted images (in which bright cluster galaxies are subtracted from the mosaic) from the UNCOVER survey DR3, since some of the sample galaxies are near one of the bright cluster galaxies. Using the non-subtracted science image can lead to contamination and cause our \statmorph and \pysersic fits to fail due to poor sky subtraction, which is driven by the high flux of a nearby bright cluster galaxy. From the UNCOVER survey DR3, we also provide the PSF (Point Spread Function of the filter), inverse-variance map, and segmentation map to \statmorph and \pysersic in 12\arcsec\ cutouts.  

Furthermore, we mask out nearby objects using the segmentation map, thereby enabling optimal sky subtraction and better galaxy fitting. In our \pysersic pipeline, we explicitly mask out pixels with invalid values in the weight maps (where weight $\leqslant 0$). These pixels are added to the mask to exclude them from the fit. As an additional safeguard, we set the RMS values for these invalid pixels to $10\times$ the median RMS of the valid pixels, ensuring they do not bias the model.

For \texttt{pysersic}, we provide all the above-mentioned inputs and fit a \texttt{sersic\_exp} model to the galaxy. The \texttt{sersic\_exp} model is a combination of a \sersic \citep[for the bulge]{Sersic1968} and an exponential profile \citep[for the disk]{Freeman1970}. We use the \texttt{autoprior} feature of \pysersic to automatically set the priors for the parameters based on the input image. This feature uses \texttt{photutils} \citep{Bradley2025} to generate priors by measuring galaxy properties within the image cutout. We manually give the X\textsubscript{c} and Y\textsubscript{c} priors, where  X\textsubscript{c} and Y\textsubscript{c} are the x and y coordinates of the galaxy centroid respectively, in a $20\times20$ (for SW filters: F090W, F115W, F150W, F200W) and $10\times10$ (for LW filters: F277W, F356W, F444W) pixel area around the galaxy center detected by \texttt{photutils} via the \texttt{autoprior} feature, to avoid a centroiding error. For some galaxies with low SNR, we manually force the X\textsubscript{c} and Y\textsubscript{c} priors to be in the center of the image in case \texttt{photutils} is not able to detect the center properly. We use \texttt{sky\_type} = ``flat'' and estimate the posterior distribution using the \texttt{svi-flow} method. \texttt{svi-flow} is a form of variational inference that uses the block neural autoregressive flow (BNAF), a normalizing flow model, to approximate the posterior distribution \citep{decao2019blockneuralautoregressiveflow}.    

\subsection{Spatially Resolved Analysis using piXedfit}

Spatially resolved analysis is an essential part of our work because it helps us understand exactly how quenching occurs within individual galaxies. Looking at a galaxy as a single, integrated object can often obscure important local details. By measuring physical properties across different regions of a galaxy, we can try to piece together its history of mass assembly and star formation. This detailed approach allows us to observationally test whether mechanisms such as inside-out quenching are actually responsible for shutting down star formation in these massive galaxies at cosmic noon.

We use the \pixedfit package \citep{Abdurrouf2021} to perform spatially resolved analysis of the selected massive quiescent galaxies. \pixedfit is a Python package designed for pixel-by-pixel SED modeling of galaxies using multi-wavelength imaging data. It enables the extraction of spatially resolved physical properties of galaxies, such as stellar mass, star formation rate, age, metallicity, and dust attenuation, by modeling the observed SEDs of individual pixels within a galaxy. It involves several key steps, which we describe below.

\subsubsection{Creating an image datacube}

Using the \texttt{piXedfit\_images} module, we begin by preparing multi-wavelength imaging data of the target galaxies, ensuring that the images are aligned and have consistent pixel scales ($0.04\arcsec$). This may involve resampling and reprojecting images from different filters to a common grid. We use the F444W filter as the reference for the PSF matching. We use the PSF matching kernels from the UNCOVER DR3 \citep{Suess2024}. We then combined the images into a datacube, with each slice corresponding to a different wavelength or filter. Also, a region of interest is defined within the datacube to focus on the target galaxy using \texttt{SEP} (Source Extractor as a library) \citep{Bertin1996, Barbary2016} via \texttt{piXedfit}'s image processing module, which can add different filters' segmentation maps together to construct the segmentation map. We used the F115W, F150W, and F200W filters to construct the segmentation map. The parameter settings for SEP are as follows:
\begin{itemize}
    \item \texttt{minarea} $= 40$, minimum number of pixels above the threshold for an object to be detected.
    \item \texttt{thresh} $= 2.0$, detection threshold in units of the background RMS.
    \item \texttt{deblend\_nthresh} $= 40$, number of thresholds used for deblending overlapping objects.
    \item \texttt{deblend\_cont} $= 0.001$, minimum contrast ratio for deblending.
\end{itemize}
We use a tight deblending contrast ratio because Abell 2744 is a dense field. Additionally, two of our sample galaxies ($\texttt{ID\_DR3}=20697$ \& 45378) contained foreground objects that were not deblended correctly and therefore, not included as separate objects in the UNCOVER catalog. These galaxies are distinctly visible in the shorter-wavelength filters but are faint in the longer-wavelength filters (see Figure~\ref{fig:rgb_collage}). We assume them as foreground galaxies and deblend them. We then run \bagpipes on these foreground galaxies. They are low mass ($\log(M_*/M_\odot)\approx 8.0$ and $7.4$ respectively) and at lower redshifts ($z \approx 1.4$ and $1.94$ respectively). In the corner plots of the galaxy near 20697, we can see that the redshift posterior has a very small bump at $z \sim 2.1$. This is similar to the redshift of our main galaxy ($z \sim 2.24$), so there is a small probability that the foreground galaxy is actually a satellite of the main galaxy. However, the redshift posterior could also be attributed to contamination from the main galaxy in the foreground galaxy's photometry, which could lead to a bimodal posterior. We would require spectroscopic data to confirm whether it is a satellite. All the \pixedfit analyses are done assuming that this object is a foreground galaxy and is therefore masked out from the main galaxy. The galaxy near 45378 has a redshift significantly lower than that of the main galaxy ($z \sim 2.3$) and a narrow redshift posterior, so it is unlikely to be a satellite. We have also masked it out of the main galaxy in our \pixedfit analysis.

\subsubsection{Pixel Binning}
For spatially resolved SED modeling, we apply pixel binning to the datacube using the \texttt{piXedfit\_bin} module. This process groups neighboring pixels based on their signal-to-noise ratio (SNR), spatial proximity, and SED similarity, forming bins that preserve spatial structure while improving SNR for more robust SED modeling. The method utilizes the Voronoi binning algorithm of \citet{Cappellari2003}, producing compact, non-overlapping bins with similar SNR in a chosen band. The algorithm has been modified by \citet{Abdurrouf2017} to also account for SED shape similarity among pixels. The SED shape similarity helps us achieve good, consistent SED modeling for all the bins. The parameters used for pixel binning are as follows:
    
\begin{itemize}
    \item \texttt{ref\_band} = F200W, index of the reference band (filter) for sorting pixels based on the brightness. The central pixel of a bin is the brightest pixel in this reference band. We use F200W as the reference band because it corresponds to the rest-frame optical wavelength for all our sample galaxies.
    \item \texttt{target\_snr} $= 5$, target signal-to-noise ratio for each bin.
    \item \texttt{Dmin\_bin} $= 8$ pixels, minimum size of each bin. This is high so that we get sufficient SNR in the F070W and F090W filters, even though we have not enforced a target SNR for these filters.
    \item \texttt{del\_r} $= 2$ pixels, it is the increment of circular radius (in units of pixels) adopted in the pixel binning process.
    \item \texttt{redc\_chi2\_limit} $= 3$, reduced chi-squared threshold below which two pixels' SEDs are considered similar in shape and are added to the bin.
\end{itemize}
We set an SNR threshold for each broadband filter following \citet{Haryana_2025}. Here, all the filters below the rest frame wavelength of 4000 \AA\, have the SNR threshold set to zero, with the rest of them having $\texttt{target\_snr}=5$. We set those filters to have a zero SNR threshold because the rest frame UV part of the SED is very faint for quiescent galaxies and thus has low SNR. On the other hand, we set the SNR threshold for all medium bands to zero, since medium bands have lower SNRs compared to the broadbands,  \added{due to their narrower bandwidth}. 

\subsubsection{SED modeling}
To fit the bins of the binned datacube, we first generate the model SEDs. This is done by using the \texttt{piXedfit\_model} module. It uses the Flexible Stellar Population Synthesis (FSPS) code \citep{Conroy2009, Conroy2010} to generate the model SEDs. For interface to the Python environment, the python-fsps \citep{Foreman-Mackey2014} package is used. After generating the model SEDs, we use the \texttt{piXedfit\_fitting} module to fit the observed SEDs of the binned datacube with the generated model SEDs. We fix the redshifts of the bins to the redshifts of the main galaxies, derived from \texttt{BAGPIPES} or the spectroscopic redshifts, where available. The modeling is done using a Bayesian approach, which allows us to derive posterior probability distributions for the physical parameters of interest. The parameters used are mentioned in the Table~\ref{tab:pixedfitparams}. We use the \added[citation corrected]{\citet{Kroupa2001}} initial mass function, Padova isochrones \citep{Girardi2000,Marigo2007,Marigo2008}, MILES stellar spectral library \citep{Sanchez-Blazquez2006,Falcon-Barroso2011}, \citet{Calzetti2000} dust law, \citet{Draine2007} dust emission, nebular emission modeling \citep{Byler2017} based on the \texttt{CLOUDY} code \citep{Ferland1998,Ferland2013} \added{and the delayed tau SFH}. We do not use the AGN component. Furthermore, we fix the ionization parameter to be $\log{U} = -3$ for the nebular emission modeling (similar to our \bagpipes settings).

Using the \pixedfit SED modeling results, we calculate the pixel-level stellar mass and SFR following \citet{Haryana_2025}. The pixel-level redistribution is done by weighting the flux of each pixel in a bin to the total flux of the bin in a reference band (F090W for SFR and F444W for stellar mass). We then compute the radial profiles of the stellar mass, SFR, sSFR, mass-weighted age, and dust attenuation (A$_V$). We extract the ellipticity and position angle of the galaxy from the \pixedfit F200W output (PSF-matched image) using SEP, and then make the radial profiles with elliptical annuli. \added{For consistency, we check if the derived results change for a double power law SFH, and we find no significant difference in the radial profile trends.} 

\begin{deluxetable*}{lcl}[ht!]
\tablecaption{\pixedfit Modeling Parameters\label{tab:pixedfitparams}}
\tablehead{
\colhead{Parameter} & \colhead{Value / Range} & \colhead{Description}
}
\setlength{\tabcolsep}{10pt}
\tabletypesize{\small}
\startdata
\cutinhead{Star Formation History (Delayed Tau Model)}
log\_age & [0.1, 0.6] & Log of age (Time since SF began) in Gyr \\
log\_tau & [-1.0, 1.5] & Log of e-folding timescale of SFH decay in Gyr \\
logzsol & [-0.2, 0.2] & Log of stellar metallicity ($Z/Z_\odot$) \\
\cutinhead{Dust Attenuation \& Emission}
dust2 & [0.0, 5.0] & ISM dust law param following \citet{Calzetti2000} (Optical depth) \\
log\_gamma & [-4.0, 0.0] & Log of dust emission parameter \\
log\_qpah & [-3.0, 1.0] & Log of PAH abundance \\
log\_umin & [-1.0, 1.39] & Log of minimum radiation field intensity \\
\cutinhead{Nebular Emission}
gas\_logu & -3.0 & Log of the gas ionization parameter \\
\enddata
\end{deluxetable*}

\section{Results} \label{sec:results}

\subsection{Morphological Analysis}

\begin{figure}
\centering
\includegraphics[width=0.8\columnwidth]{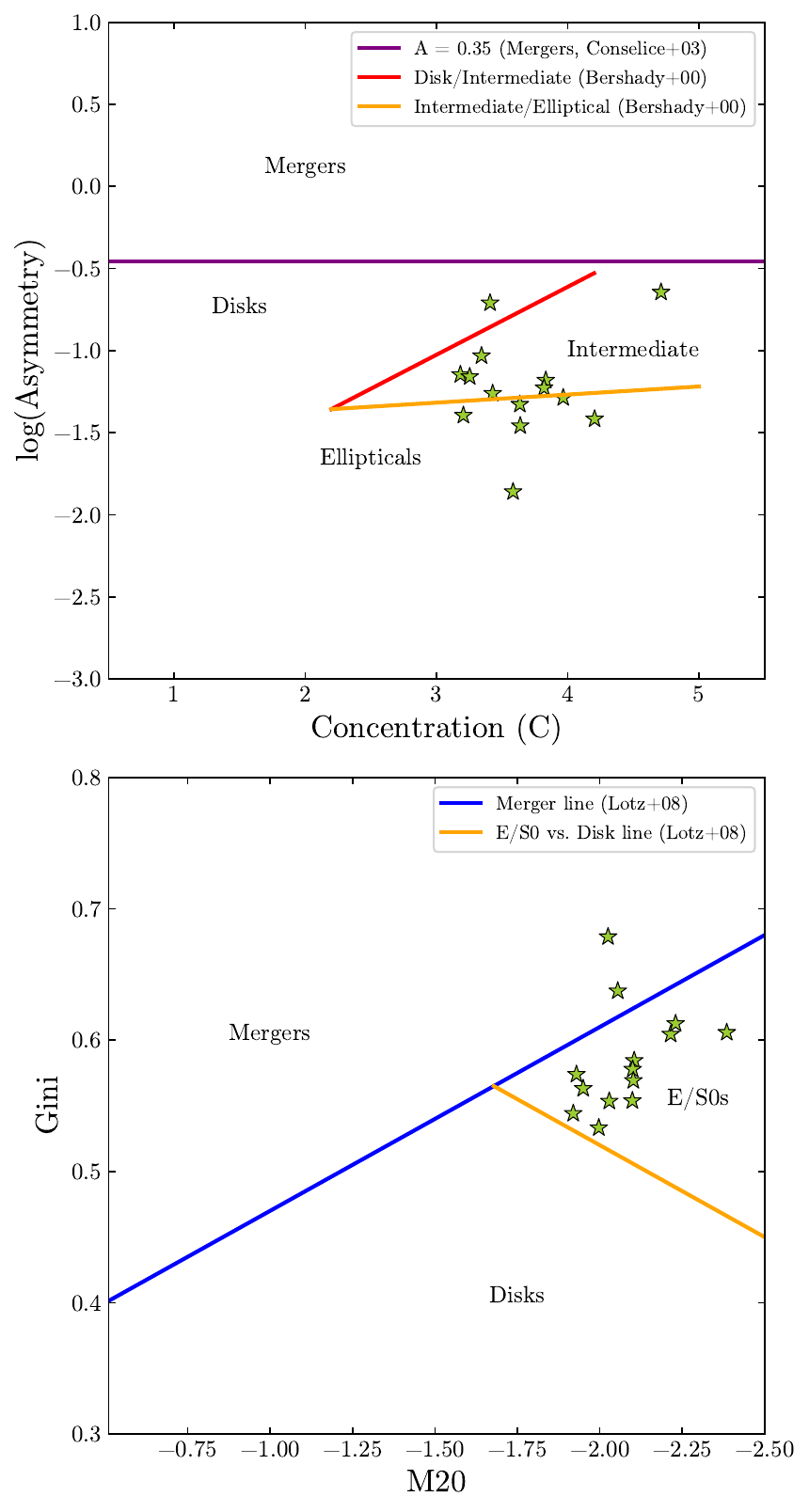}
\caption{Non-parametric morphological parameters for our sample of 14 unique massive quiescent galaxies for the F200W NIRCam filter. Top: $\log_{10}(A)$ (asymmetry) vs.  The merger line was taken from \citet{Conselice2003}, and the disk/intermediate and intermediate/elliptical boundaries were implemented following \citet{Bershady2000}. Bottom: Gini coefficient vs.\ $M_{20}$. The boundary lines were implemented following \citet{Lotz2008}.} 
\label{fig:casgm}
\end{figure}

\begin{figure*}
\centering
\includegraphics[width=\textwidth]{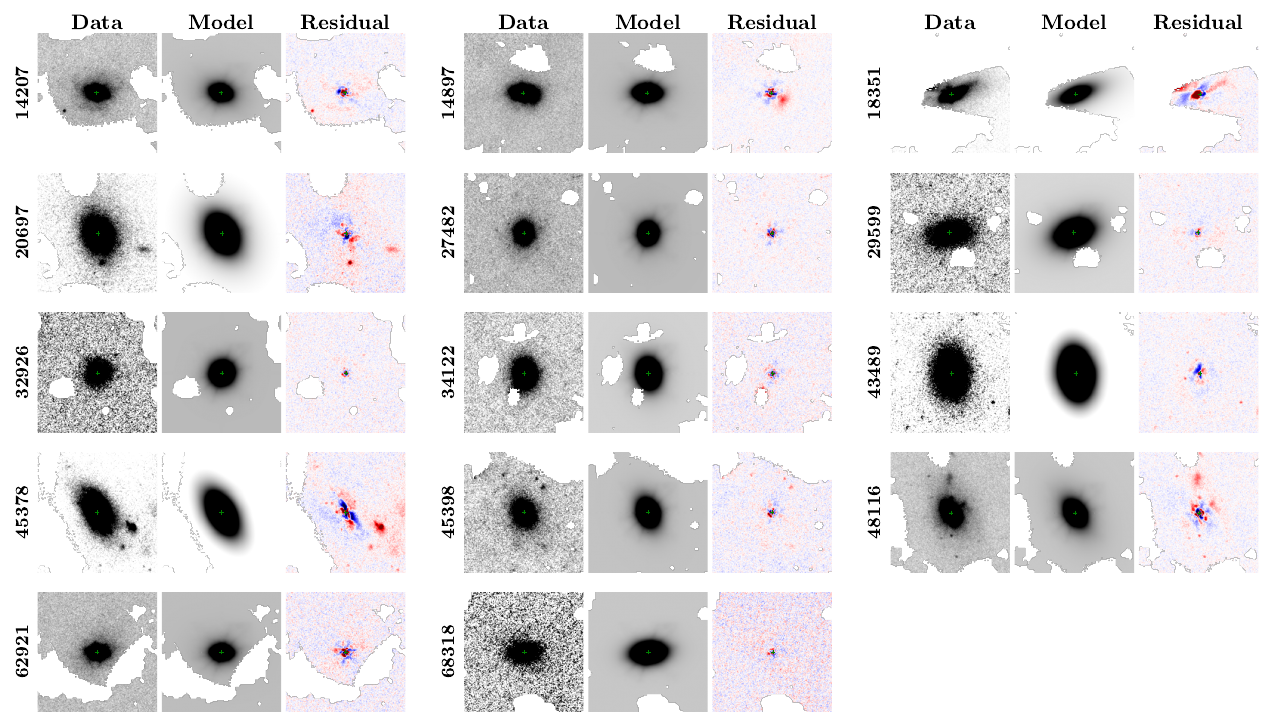}
\caption{ Two-component \pysersic fits for the galaxies in our sample in the F200W filter (4\arcsec\ cutouts). The panels display the original images, the median models, and the resulting residual maps. The color scale limits (vmin and vmax) for the residual plots are set to $\pm 0.1$ (in $10\ \mathrm{nJy}$), where blue regions indicate less flux than the model and red regions indicate excess flux. The average absolute residual is $4.16\% \pm 1.39\%$, with values ranging from 0.38\% to 22.09\%. The standard deviation of the mean is calculated via bootstrapping for 1000 iterations.}   
\label{fig:F200W_pysersic}
\end{figure*}

\begin{figure*}
\centering
\includegraphics[width=\textwidth]{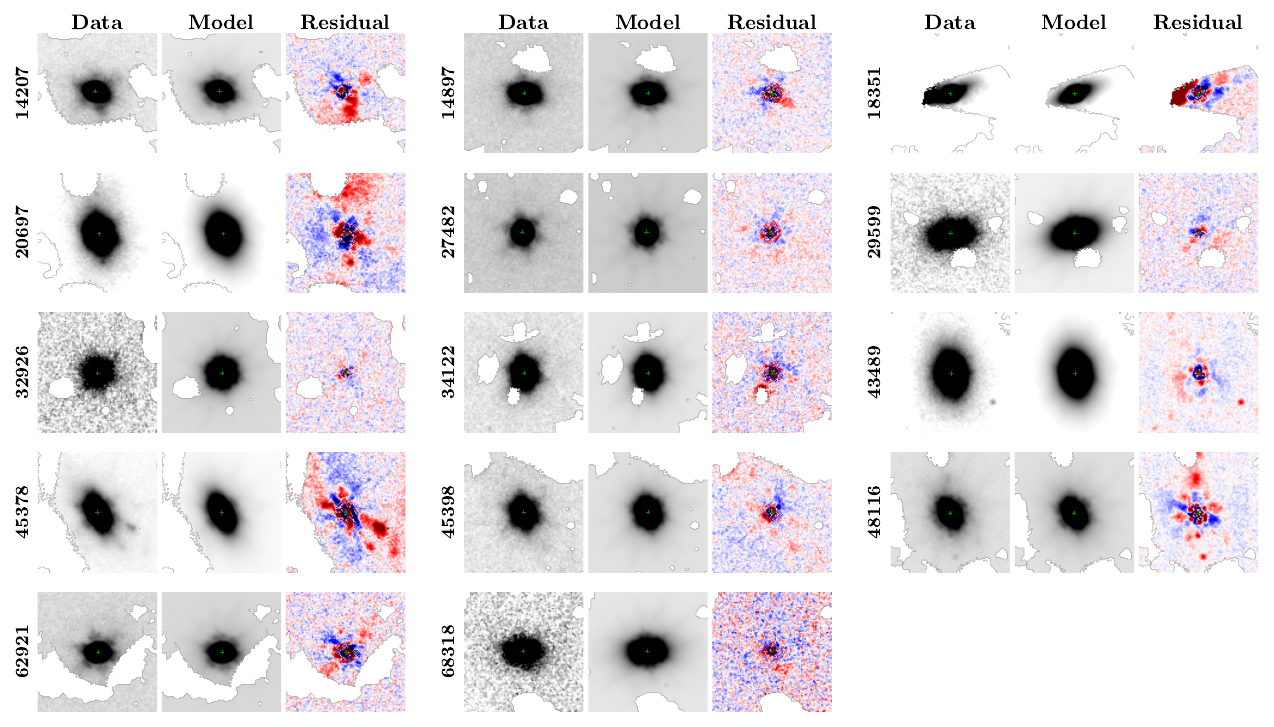}
\caption{Two-component \pysersic fits for the galaxies in our sample in the F444W filter (4\arcsec\ cutouts). The panels display the original images, the best-fit models, and the resulting residual maps. The color scale limits (vmin and vmax) for the residual plots are set to $\pm 0.1$ (in $10\ \mathrm{nJy}$), where blue regions indicate less flux than the model and red regions indicate excess flux. The average absolute residual is $1.48\% \pm 0.67\%$, with values ranging from 0.16\% to 10.25\%. The standard deviation of the mean is calculated via bootstrapping for 1000 iterations.}   
\label{fig:F444W_pysersic}
\end{figure*}

We utilize non-parametric statistics, specifically the Concentration ($C$) versus Asymmetry ($\log A$) and Gini versus $M_{20}$ diagrams, to classify them. As defined by \citet{Conselice2003}, elliptical galaxies typically have smooth, concentrated light profiles, placing them in regions of high $C$ and low $A$. Disk galaxies exhibit lower concentration, while mergers are identified by very high asymmetry ($A > 0.35$). In the Gini--$M_{20}$ plane \citep{Lotz2004, Lotz2008}, early-type galaxies (E/S0/Sa) reside in regions of high Gini coefficient and low $M_{20}$, whereas late-type disks show lower Gini coefficients, and mergers occupy the parameter space with highly positive $M_{20}$ values or exceptionally high Gini coefficients. Most galaxies in our sample fall into the Intermediate/S0s or Elliptical categories. On the other hand, our sample galaxies possess a significant bulge component, as evidenced by high \sersic indices ($n$) and Bulge to Total luminosity ratios ($B/T$) derived from our \pysersic fits (with median $n$ ranging from $1.73_{-0.39}^{+0.29}$ in F115W to $4.25_{-0.47}^{+1.18}$ in F444W, and median $B/T$ ranging from $0.54_{-0.11}^{+0.07}$ in F115W to $0.76_{-0.06}^{+0.02}$ in F444W). The morphological parameters for all the galaxies are presented in Appendix~\ref{sec:morph_results}, and the \statmorph results for the F200W filter are shown in Figure~\ref{fig:casgm}. They include the disk/intermediate and the intermediate/Elliptical boundaries from \citet{Bershady2000} in the $\log(A)$ vs. $C$ diagram. The \pysersic results for the F200W and F444W filters are shown in Figure~\ref{fig:F200W_pysersic} and Figure~\ref{fig:F444W_pysersic}, respectively. 

A significant subset of our sample (\texttt{ID\_DR3} = 27482, 29599, 32926, 45398) represents the ``standard'' massive quiescent galaxy archetype. They are quite isolated, possess smooth and undisturbed morphologies, and securely reside in the Intermediate/Elliptical parameter space in almost all filters. Their \pysersic fits yield very low residuals, and they are overwhelmingly bulge-dominated. \texttt{ID\_DR3} = 68318 (the faintest source in our sample) similarly maintains an elliptical categorization, though its lower signal-to-noise ratio introduces more scatter in its parametric (\texttt{pysersic}) fits.

\begin{figure}
\centering
\includegraphics[width=0.8\columnwidth]{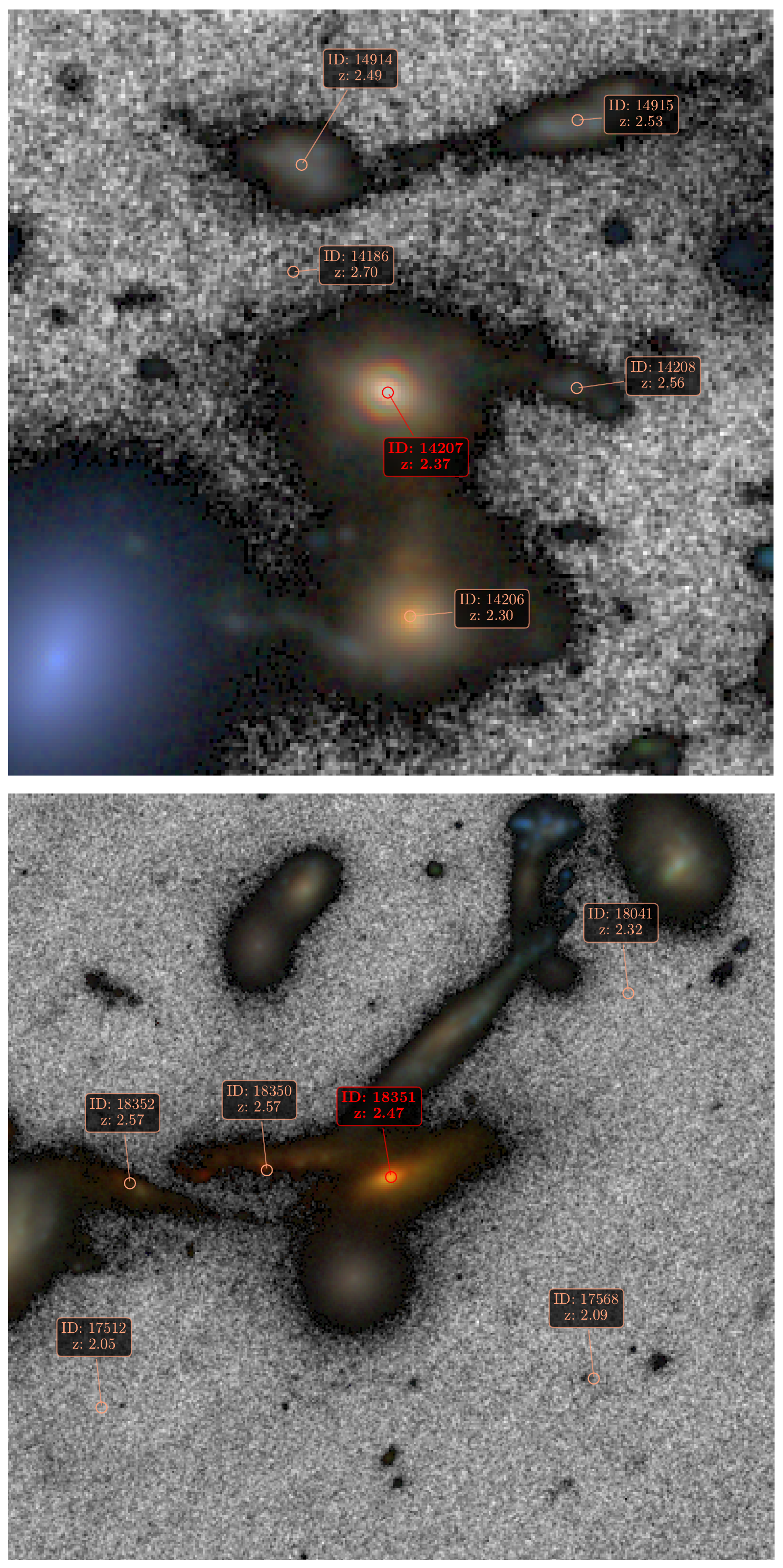}
\caption{Top: This is an RGB image (using F444W, F356W, and F277W) of galaxy \texttt{ID\_DR3} = 14207, which is likely to have undergone tidal disruptions due to interactions with \texttt{ID\_DR3} = 14206, distinctly visible in the long wavelength filters. Bottom: This is an RGB image (using F200W, F150W, and F115W) of galaxy \texttt{ID\_DR3} = 18351, which is likely to be a merger candidate. We can see nearby objects to its left with very similar redshifts. Both the cutouts are of 8\arcsec. Surrounding galaxies with similar redshifts (obtained using \texttt{BAGPIPES}, within the error bars of their photometric redshifts) are marked in the cutouts.} 
\label{fig:rgb_18351}
\end{figure}

The galaxies \texttt{ID\_DR3} = 14207, 14897, and 18351 demonstrate strong signs of interactions or ongoing mergers, two of them are shown in Figure~\ref{fig:rgb_18351}. They often sit close to or within the merger regions of the $C$-$\log A$ and Gini-$M_{20}$ diagrams at shorter wavelengths. \texttt{ID\_DR3} = 14207 shows signs of tidal disruption with a distinct tail visible in its long-wavelength residuals (see Figure~\ref{fig:F444W_pysersic}). \texttt{ID\_DR3} = 14897 is a compelling case of an ongoing merger, featuring two distinct cores that shift its classification toward merger and disk-dominated in short-wavelength filters (F090W and F115W) while remaining bulge-dominated at longer wavelengths. \texttt{ID\_DR3} = 18351 is highly disturbed with several nearby objects at similar redshifts (see Figure~\ref{fig:rgb_18351}), resulting in predictably large \sersic fitting residuals across all filters. 

Several galaxies (\texttt{ID\_DR3} = 34122, 43489, 48116, 62921) exhibit complex, wavelength-dependent morphologies. They are generally classified as Intermediate/Elliptical. They are disk-dominated in the shortest wavelength filters but are bulge-dominated at longer wavelengths. \texttt{ID\_DR3} = 48116, in particular, displays a distinctly visible star-forming clump slightly north of its center, leading to clumpy \pysersic residuals. 

\added[Ref. comment: soften AGN claim]{Inspection of the \pysersic residuals occasionally reveals very faint central structures or alternating positive/negative residuals at the cores of 6 galaxies (\texttt{ID\_DR3} = 14207, 14897, 27482, 43489, 48116, and 62921). While these features could in principle point to a compact central component such as a weak active galactic nucleus (AGN), they are very small and could also easily arise from systematic modeling limitations, such as small PSF mismatches, lensing-induced shear, or deblending errors.} 

Two galaxies, \texttt{ID\_DR3} = 20697 and 45378, highlight edge cases caused by observational limitations. Both have nearby foreground galaxies that the UNCOVER DR3 segmentation map did not deblend. This directly affected their \statmorph classifications, erroneously pushing them into the merger categories at shorter wavelengths where the foreground objects are relatively bright. Fortunately, our two-component \pysersic modeling isolated the separate structures by excluding the foreground objects, resulting in clear detections in the residuals and reliable structural parameters for the target galaxies, confirming they are inherently bulge-dominated.

\subsection{piXedfit Analysis}

Using \texttt{piXedfit}, we reconstructed spatially resolved maps of the stellar mass and SFR, and sSFR across the sample. Spatially resolved properties allow us to cleanly decouple localized structures, determine whether a galaxy is undergoing inside-out or outside-in quenching, and trace minor mergers or residual star-forming regions. Grouping our targets by their resolved spatial characteristics reveals several defining themes across the sample. 

The vast majority of our sample (\texttt{ID\_DR3} = 14207, 14897, 18351, 20697, 27482, 29599, 32926, 34122, 45398, 62921, 68318) exhibits a consistently positive sSFR gradient. This indicates that their inner cores are quenching faster than their outskirts, strongly supporting the inside-out quenching scenario. Within this group, smooth isolated galaxies (e.g., \texttt{ID\_DR3} = 27482, 29599, 45398, and 62921) have a radial SFR profile that decreases initially but flattens or rises slightly past $\sim 2-3.5$~kpc, highlighting residual low-level star formation in their outer disks.

Three galaxies (\texttt{ID\_DR3} = 43489, 45378, 48116) demonstrate ambiguous or fluctuating quenching gradients, where the sSFR visibly decreases and then increases farther out. However, their pixel-level SFR and stellar mass maps reveal a common physical driver in the form of secondary, gas-rich star-forming clumps or minor mergers. \texttt{ID\_DR3} = 43489 has a secondary star-forming core towards the west of its center with $\sim 2$~dex lower stellar mass than the central core. Similarly, \texttt{ID\_DR3} = 45378 and 48116 both feature distinctly visible off-center star-forming clumps in their SFR maps.

The spatially resolved maps also uniquely illuminate the gas and mass dynamics across our major merger candidates. For \texttt{ID\_DR3} = 14207, the tidal disruption visible in the broadband images is distinctly recovered in the pixel-level stellar mass map. Interestingly, this structure is notably absent in the SFR map, suggesting the tidal interaction has primarily perturbed older stars without triggering a gas-rich starburst. On the other hand, \texttt{ID\_DR3} = 14897 shows a distinctly measurable bump in its radial SFR profile at $\sim 2$~kpc directly corresponding to the secondary star-forming core. Finally, \texttt{ID\_DR3} = 18351 shows a highly irregular SFR map and a region of high stellar mass density and dust extinction towards the left of the center, firmly establishing it as a dusty, ongoing merger with ongoing star formation.

\begin{figure*}
\centering
\includegraphics[width=0.9\textwidth]{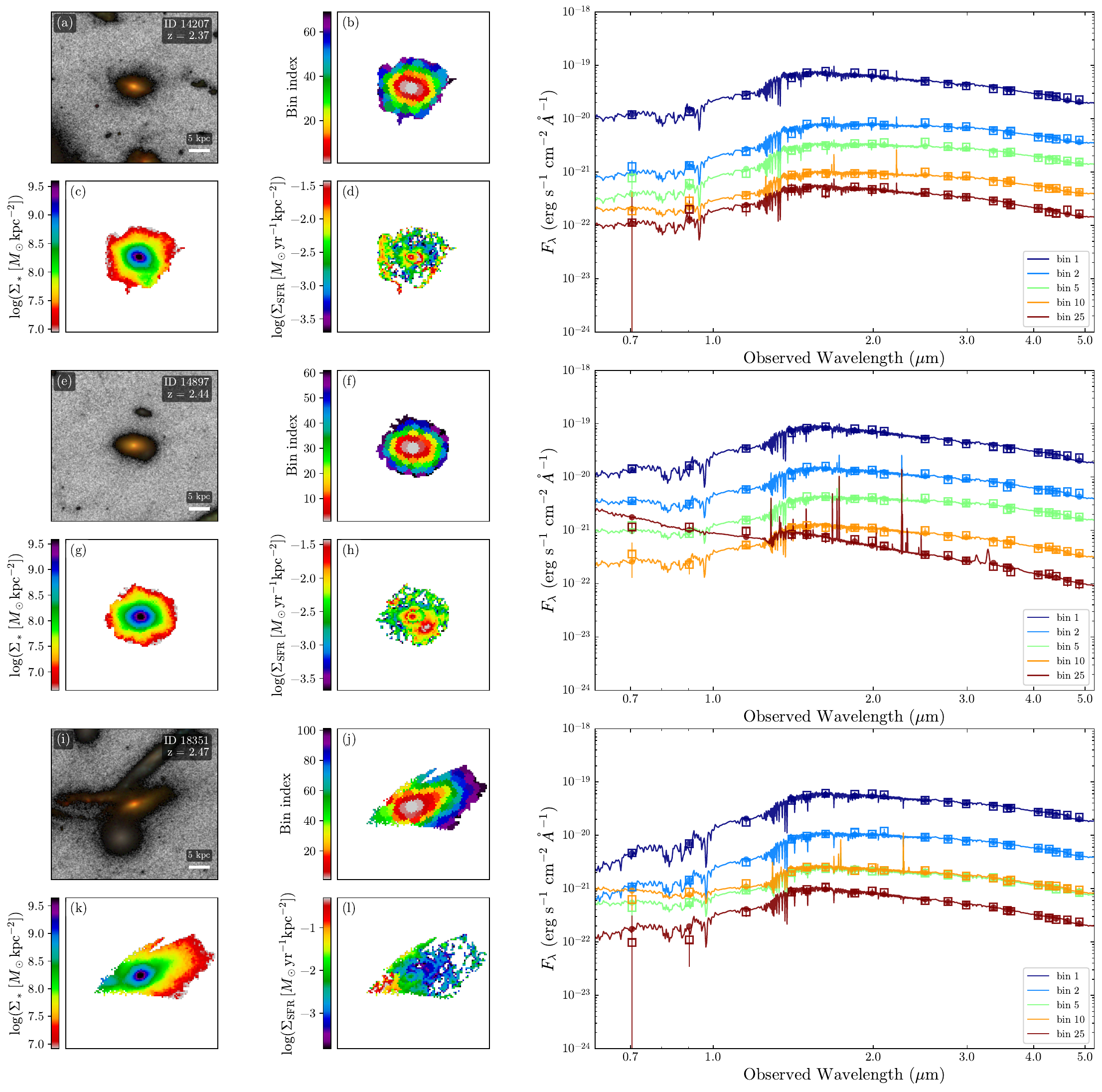}
\caption{Figures (a), (e), and (i) are the RGB images (using F115W, F150W, and F200W filters), (b), (f), and (j) are the bin index maps, (c), (g), and (k) are the pixel-level stellar mass maps, and (d), (h), and (l) are the pixel-level SFR maps for the three interacting galaxies in our sample (\texttt{ID\_DR3} = 14207, 14897, and 18351). The spectra of different bins for these three galaxies are shown in the figures on the right. Filled circles are the model fluxes, and the open squares are the observed fluxes, with error bars. The solid lines are the best-fit SEDs for each bin.}
\label{fig:haryana_style_figure}
\end{figure*}

Our \pixedfit results for the three interacting galaxies in our sample are showcased in Figure~\ref{fig:haryana_style_figure}.

\subsection{Radial Profiles}

\begin{figure*}
\centering
\includegraphics[width=0.85\textwidth]{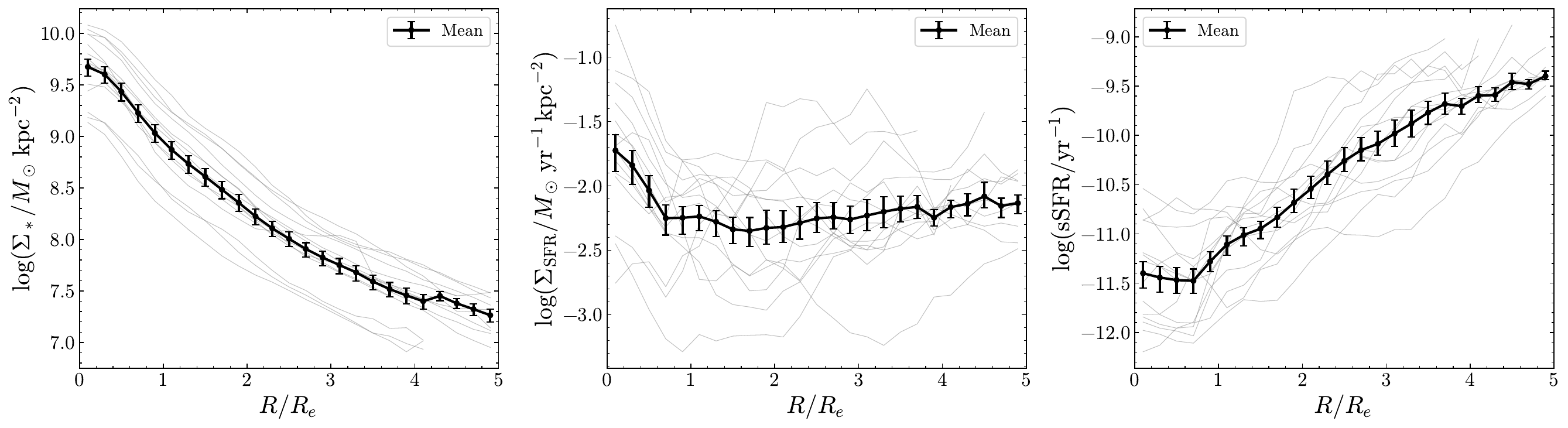}
\includegraphics[width=0.85\textwidth]{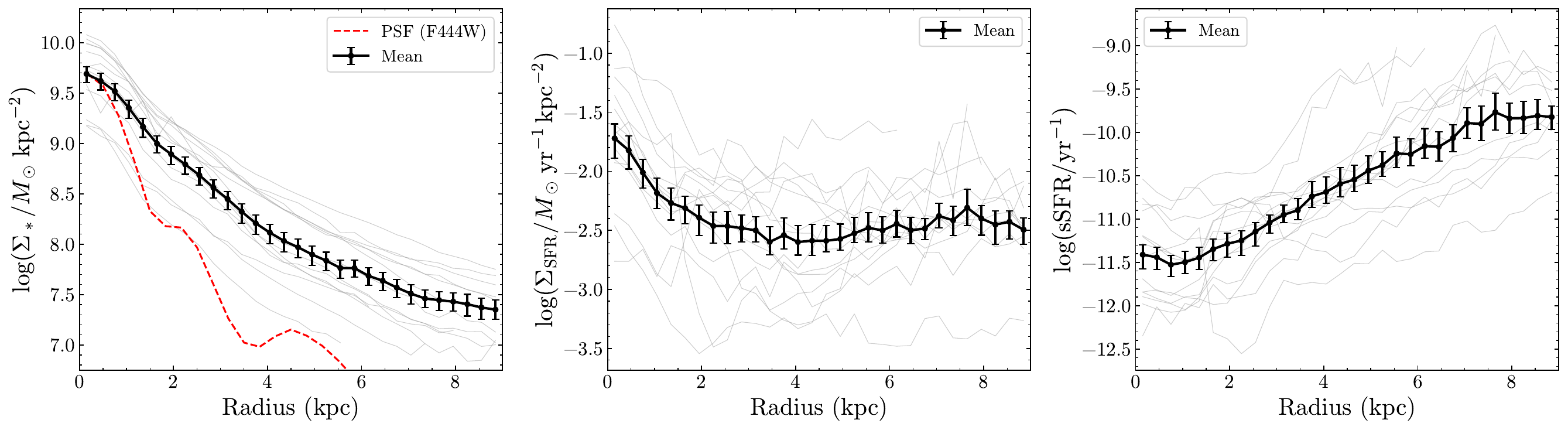}
\caption{Mean radial profiles for the stellar mass, SFR, and sSFR for our sample of 14 galaxies. The top panels show the profiles normalized by the half mass radius ($R_e$), following \citet{Laishram2025}. The bottom panels show the profiles with the radius $R$ in increments of 0.3 kpc for direct comparison with \citet{Haryana_2025}. Left: Stellar mass radial profiles. Middle: SFR radial profiles. Right: sSFR radial profiles. The red dashed line in the stellar mass radial profile plot in the bottom panel is the PSF profile for the F444W filter, which is the reference filter for the PSF matching. We can observe the increasing sSFR gradient, which indicates inside-out quenching. The error bars are the 68\% confidence intervals of the mean, derived from bootstrapping for 1000 iterations.}
\label{fig:radial_profiles}
\end{figure*}

We plot the mean radial profiles for the stellar mass, SFR, and sSFR for the 14 galaxies in our sample in Figure~\ref{fig:radial_profiles}. We follow the method of \citet{Laishram2025} to plot the mean radial profiles with the 68\% confidence interval of the mean, derived from bootstrapping for 1000 iterations (top panel). This uses the half-mass radius ($R_e$) of each galaxy to normalize the radial profiles. The bins are in increments of 0.2 $R/R_e$.  We also plot the mean radial profiles but with the $R$ in 0.3 kpc increments instead of $R/R_e$ (bottom panel), to enable a direct comparison with the radial profiles in \citet{Haryana_2025}. The overall trend matches the inside-out quenching scenario. The SFR radial profile decreases by about 1 dex near the center, then remains constant farther out. This implies that the cores of the galaxies have high star formation rates, but the decrease in stellar mass with increasing radius is faster than the decrease in SFR, leading to a positive sSFR gradient. 

\begin{figure}
\centering
\includegraphics[width=0.8\columnwidth]{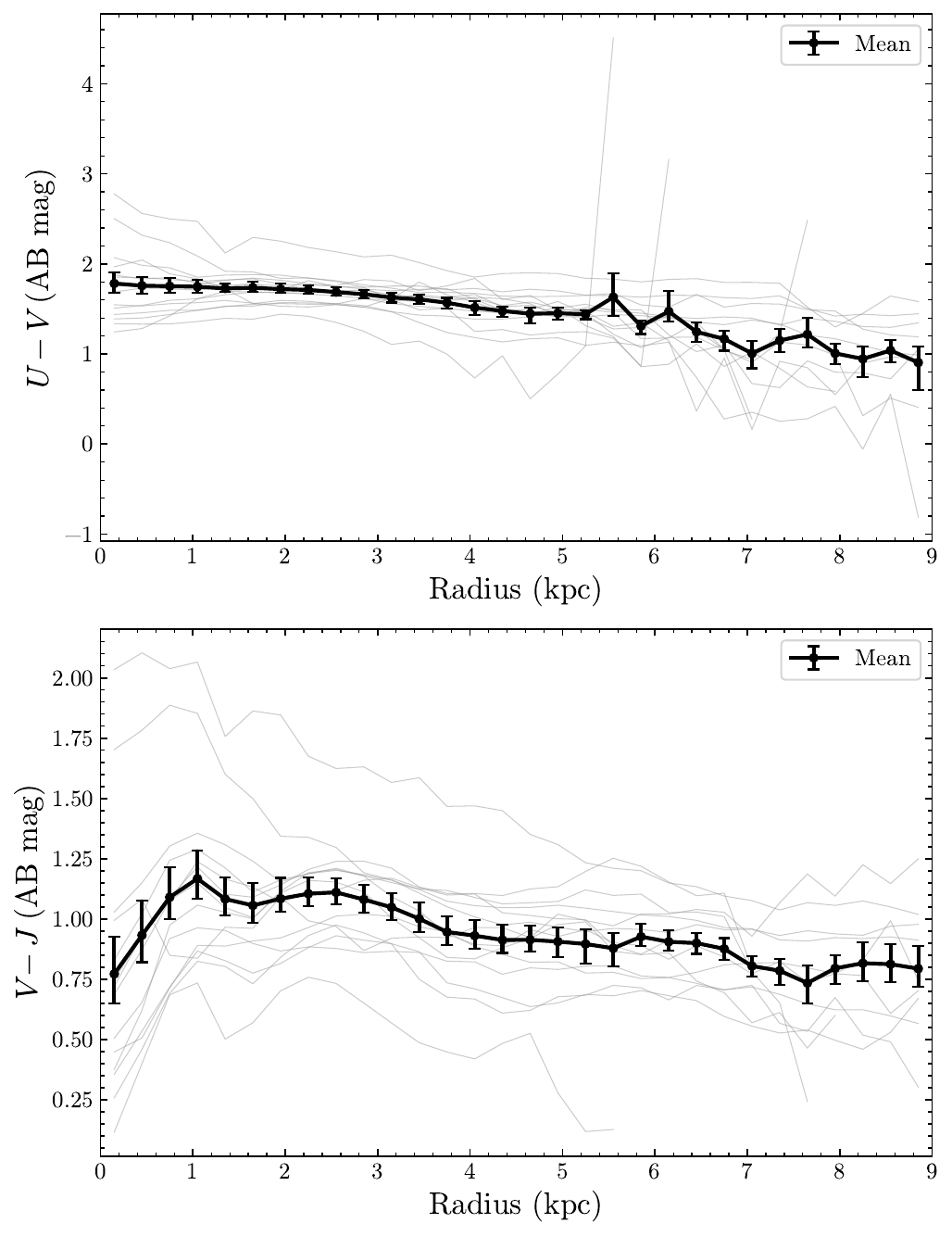}
\caption{Mean radial profiles of the $U-V$ and $V-J$ colors as a function of radius in kpc for our sample of 14 galaxies. The error bars are the 68\% confidence intervals of the mean, derived from bootstrapping for 1000 iterations. The profiles are derived from the \pixedfit SED modeling.}
\label{fig:uvj_radial_profiles}
\end{figure}

Additionally, we calculate the radial $U-V$ and $V-J$ radial profiles following \citet{Haryana_2025}, shown in Figure~\ref{fig:uvj_radial_profiles}. We define the $U$ and $V$ magnitudes using the filter curves computed by \citet{Maiz-Apellaniz2006}, whereas we compute the $J$ magnitude using the Two Micron All Sky Survey J-band transmission curve \citep{Skrutskie2006}. We derive the flux contribution at the pixel level by distributing the modeled $U$, $V$, and $J$ fluxes of each bin to individual pixels. We scale them according to the observed flux in each pixel at the rest-frame wavelength of the broadband filter closest to the target band, normalized by the total observed flux within the bin. 

\begin{figure}
\centering
\includegraphics[width=1\columnwidth]{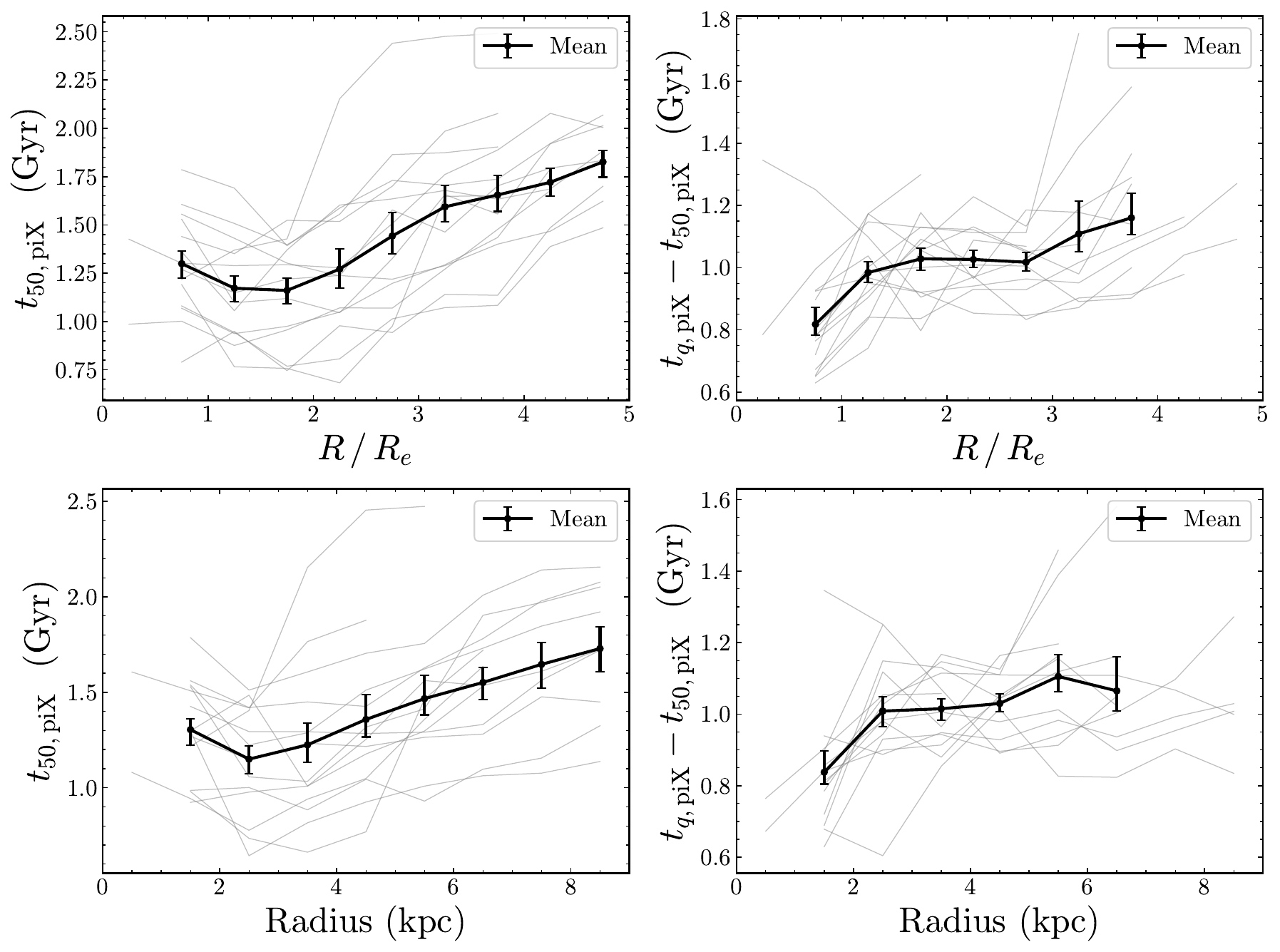}
\caption{Mean radial profiles of the formation time ($t_{50, \,\mathrm{piX}}$, on the left) and the quenching timescale ($t_{q, \,\mathrm{piX}}- t_{50, \,\mathrm{piX}}$, on the right) for our sample of 14 galaxies. The top panels show the profile normalized by the half mass radius ($R_e$) in bins of 0.5 $R_e$, while the bottom panels show the profile with radius in kpc in bins of 1 kpc. The error bars are the 68\% confidence intervals of the mean, derived from bootstrapping for 1000 iterations. The positive gradient of the formation time profile indicates that the inner regions ($<4$ kpc) of the galaxies formed earlier than the outer regions ($>4$ kpc) by $\approx 0.5$ Gyr, whereas the quenching timescale profile implies that the cores were quenched faster than the outer regions.}
\label{fig:tau_radial_profiles}
\end{figure}

\begin{figure*}
\centering
\includegraphics[width=\textwidth]{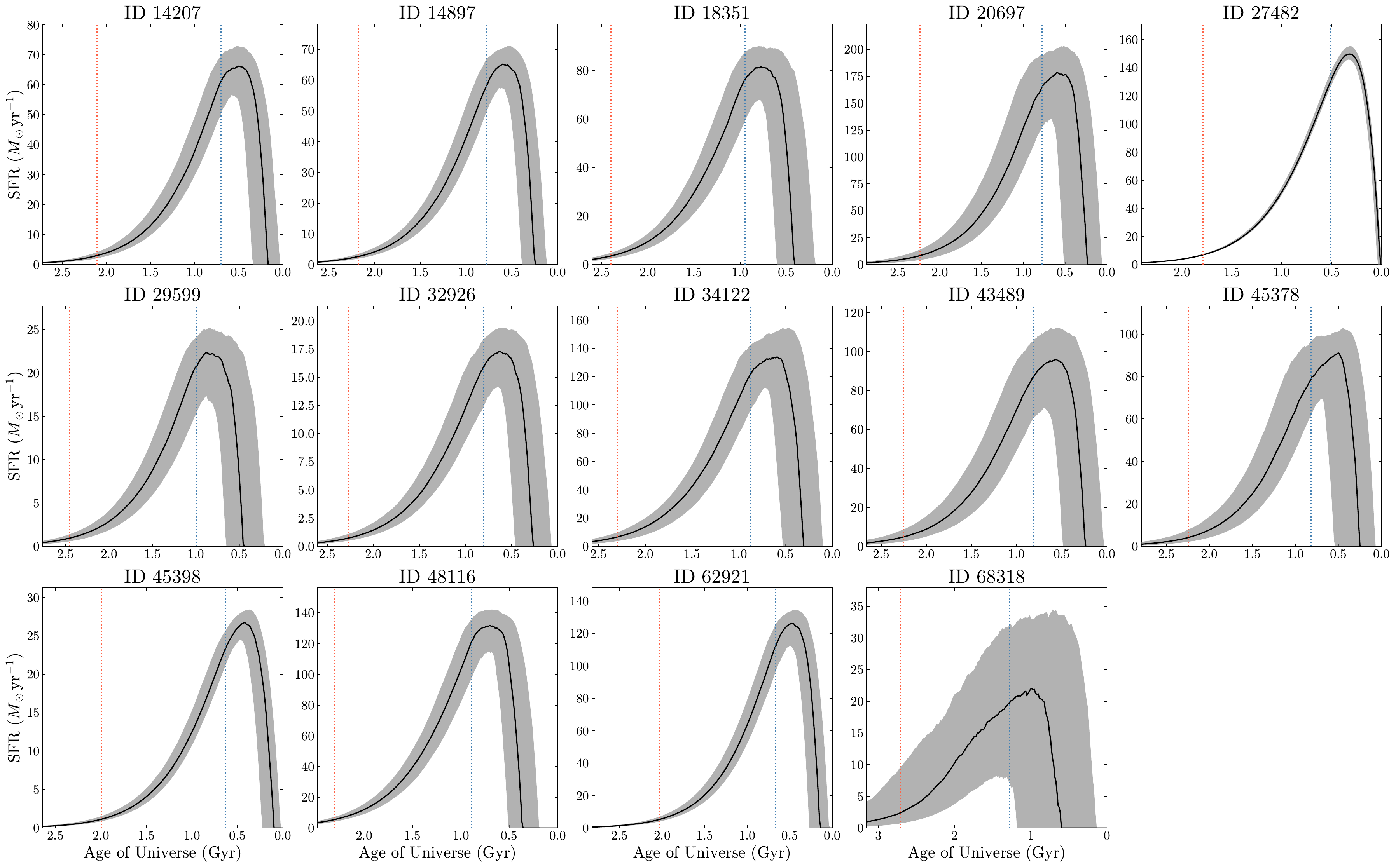}
\caption{Star formation histories (SFHs) for all galaxies in our sample from \bagpipes SED modeling. Each panel shows the posterior SFH for one galaxy, with the blue vertical line indicating the formation time ($t_{50,\, \mathrm{BG}}$) and the red vertical line indicating the quenching time ($t_{q,\, \mathrm{BG}}$). The SFHs indicate rapid early mass assembly followed by quenching in $\approx 1.4$ Gyr.}
\label{fig:sfh_grid}
\end{figure*}

Lastly, we compute the formation time ($t_{50,\, \mathrm{piX}}$) and the quenching timescale ($t_{q, \,\mathrm{piX}} - t_{50,\, \mathrm{piX}}$) radial profiles. $t_{50, \,\mathrm{piX}}$ is the age of the universe when the galaxy region reaches half its total mass, whereas $t_{q, \,\mathrm{piX}}$ is the age when its current SFR reaches 10\% of its time-averaged SFR. The $t_{50, \,\mathrm{piX}}$ radial profiles also show a positive gradient, indicating that the inner regions ($<4$ kpc) of the galaxies formed earlier than the outer regions ($>4$ kpc) by $\approx 0.5$ Gyr. The quenching timescale radial profiles imply that the cores of the galaxies were quenched more rapidly than the outskirts (see Figure~\ref{fig:tau_radial_profiles}). We also plot the Star Formation Histories (SFHs) of all the galaxies in our sample from \bagpipes SED modeling in Figure~\ref{fig:sfh_grid}. The figure also shows the quenching timescales derived from \texttt{BAGPIPES}, with the sample exhibiting a mean quenching timescale of $\Delta t_{BG} \approx 1.4$ Gyr. \added[claim removed from line 805]{To check for consistency in the radial profiles across different lensed images of \texttt{ID\_DR3} = 45378 and 45398, we ran \pixedfit on the other lensed images as well and found no significant change in the sSFR, $t_{50}$, and quenching timescale radial profile trends.} 

All the radial profile trends are similar to those reported in \citet{Haryana_2025} for their sample of 45 galaxies in the redshift range $2 < z < 3$. The mean half-mass radius of our sample is $R_e = 1.95\pm0.13$ kpc (standard error derived via bootstrapping for 1000 iterations).

\section{Discussion} \label{sec:discussion}

\begin{figure*}[t!]
\centering
\includegraphics[width=\textwidth]{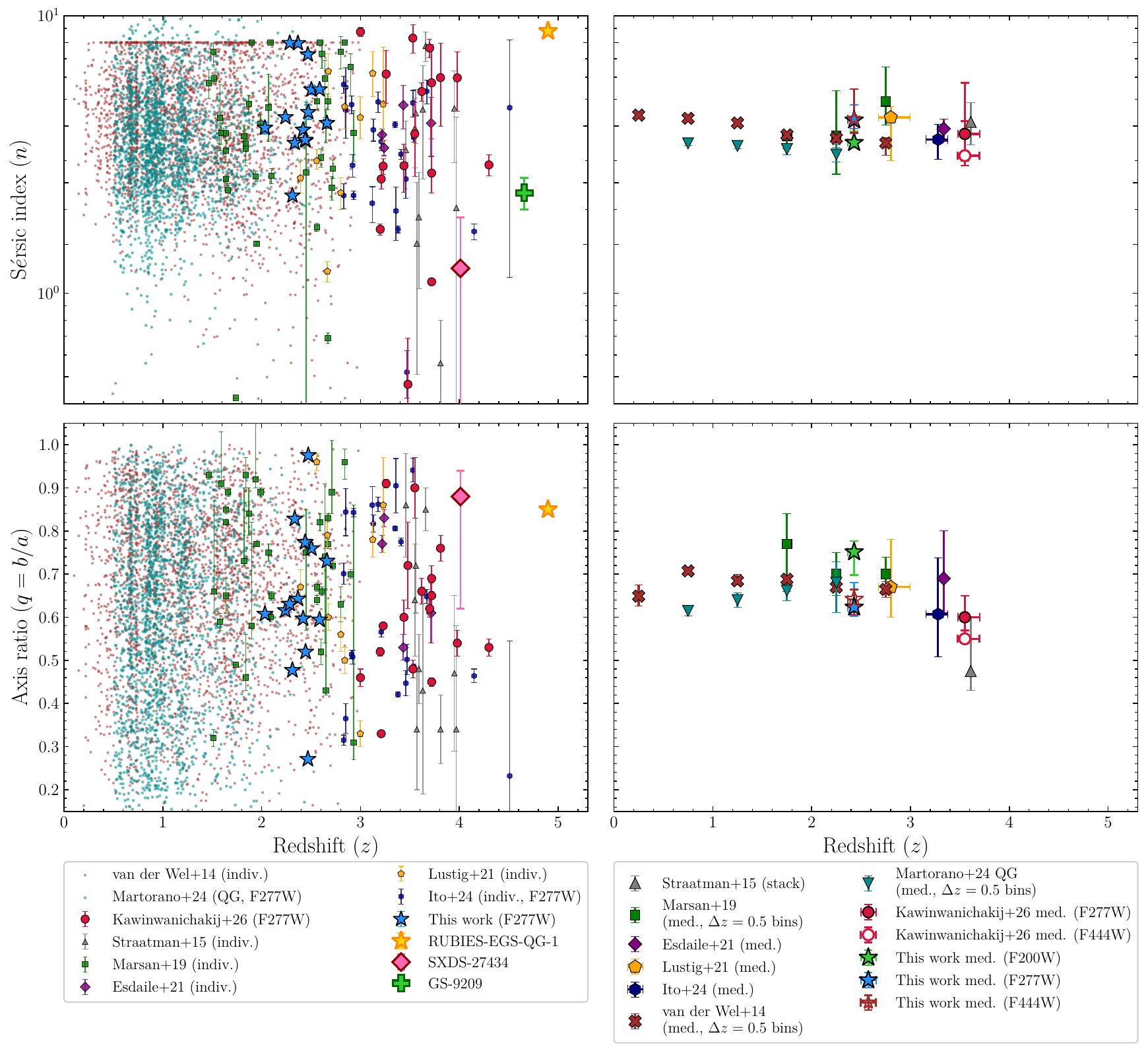}
\caption{Morphological parameters as a function of redshift for our sample of massive quiescent galaxies (F277W, blue filled stars) compared with literature values. Top panels: \sersic index ($n$) comparison. Bottom panels: Axis ratio ($q$) comparison. Comparison samples include UVJ-selected quiescent galaxies from \citet[$\log(M_*/M_\odot) > 10$]{vanderWel2014}, \citet[$10.6 < \log(M_*/M_\odot) < 11.25$]{Straatman2015}, and \citet[$9.8 < \log(M_*/M_\odot) < 11.4$]{Ito2024}; photometrically selected galaxies from \citet[$\log(M_*/M_\odot) > 11.25$]{Marsan2019}; and sSFR-selected quiescent galaxies from \citet[$\log(M_*/M_\odot) > 10$]{Martorano2024} (F277W morphology cross-matched with COSMOS2020 catalog; \citealt{Weaver2022}). Individual spectroscopically confirmed massive quiescent galaxies are also shown: SXDS-27434 \citep[$\log(M_*/M_\odot) = 11.06$]{Tanaka2019,Valentino2020,Ito2024}, GS-9209 \citep[$\log(M_*/M_\odot) = 10.58$]{Carnall2023b}, RUBIES-EGS-QG-1 \citep[$\log(M_*/M_\odot) = 10.9$]{deGraaff2025}, and galaxies from \citet[$11 < \log(M_*/M_\odot) < 11.3$]{Esdaile2021}, \citet[$10.8 < \log(M_*/M_\odot) < 11.3$]{Lustig2021} and \citet[$10.2 < \log(M_*/M_\odot) < 11.2$]{Kawinwanichakij2026}. On the right panels, the median \sersic indices (top) and median axis ratios (bottom) for our sample are shown as blue-filled stars (F277W), brown-outlined stars (F444W), and green-filled stars (F200W). All the median error bars are the 68\% confidence intervals derived from bootstrapping for 1000 iterations.}
\label{fig:sersic_q_comparison}
\end{figure*}

In this work, we have carried out morphological analysis of our sample of massive quiescent galaxies using \statmorph and \texttt{pysersic}. Most sample galaxies are Intermediate-type or S0s. They are also bulge-dominated in most of the filters. The \sersic index parameterizes the galaxy's radial surface brightness profile, measuring its degree of central light concentration. While an index of $n \sim 1$ represents an exponential light profile, characteristic of a disk, higher values approaching $n \sim 4$ follow the de Vaucouleurs profile that is characteristic of classical, dispersion-dominated bulges. We plot the \sersic indices ($n$) and the axis ratio ($q$) data from the literature in Figure~\ref{fig:sersic_q_comparison}, to compare with our results. We also include the F200W filter for comparison since it is closest to the rest frame optical wavelength at redshift $\sim$ 2--3. The median values of the \sersic index and the axis ratio for our sample for various filters are given in Table~\ref{tab:median_n_q}. We note that 3 of our galaxies hit the upper limit of the \sersic index in the \pysersic fits in some filters (\texttt{ID\_DR3} = 14207, 18351, and 62921, see Table~\ref{tab:pysersic}). Also, galaxies \texttt{ID\_DR3} = 18351 and 14897 have a lower axis ratio than the rest of the sample in some filters. This could be because of the possible ongoing mergers. 

To contextualize our findings, the figures also include $UVJ$-selected quiescent galaxies from \citet{vanderWel2014,Straatman2015,Ito2024}, photometrically selected galaxies from \citet{Marsan2019}, and sSFR-selected quiescent galaxies from \citet{Martorano2024} for comparison across a wide redshift range. We obtained F277W $n$ and $q$ values from \citet{Martorano2024} and cross-matched them with the COSMOS2020 catalog \citep{Weaver2022} to obtain their sSFRs. Then we applied the same sSFR selection criteria (see Equation~\ref{eq:ssfr_cut}) to the crossmatched sample. We also include the $n$ and $q$ (where available) of spectroscopically confirmed massive quiescent galaxies SXDS-27434 \citep{Tanaka2019,Valentino2020,Ito2024}, GS-9209 \citep{Carnall2023b}, RUBIES-EGS-QG-1 \citep{deGraaff2025}, and galaxies from \citet{Esdaile2021,Lustig2021,Kawinwanichakij2026}.

We observe that the individual points have a lot of scatter, but the median values of the \sersic indices are around 4 for a wide range of redshift (1.5 to 4) as seen from Figure~\ref{fig:sersic_q_comparison}. The scatter in the $n$ and $q$ values could be attributed to the different selection criteria used in different works, and also to the different wavelengths at which the morphological analysis is done. Furthermore, our samples have axis ratios within error bars of the median values from other works at similar redshifts \citep{vanderWel2014,Marsan2019,Lustig2021,Martorano2024}. This relatively constant $n\sim 4$ for a wide range of redshifts implies that the massive galaxies have a significant classical bulge component even at high redshifts \citep{Kawinwanichakij2026}. 

This is an interesting result because it shows that the morphology (especially the bulge component) of a massive galaxy is linked to its quenching (and hence its sSFR) for a wide range of redshifts. \citet{Bait2017} found that the morphology of a massive galaxy is strongly correlated with its sSFR, regardless of its environment, in the nearby universe. Our result supports the idea that the morphology--sSFR link is present even at high redshifts. 

Transitioning to our spatially resolved analysis, from the \pixedfit results, we find that 11 out of the 14 ($\approx 79\%$) galaxies show a positive sSFR gradient, indicating that they are quenched inside-out. The remaining 3 galaxies do not have a prominent positive or negative sSFR gradient, from which we cannot definitively say if they are quenched inside-out or outside-in. Additionally, the mean sSFR radial profiles clearly point to an inside-out quenching scenario, implying the suppression of star formation in the cores of the galaxies while the outskirts remain star-forming. The mean sSFR increases from $3.40 \times 10^{-12}$ /yr at $R/R_e = 0.5$ to $3.46 \times 10^{-10}$ /yr at $R/R_e = 4.5$. This is consistent with the findings of \citet{Lin2019}, in which they found that the fraction of massive galaxies with inside-out quenching is high irrespective of the environment in our nearby universe.

The inside-out quenching might be driven by the suppression of star formation in the cores of the galaxies due to AGN feedback \citep{Bower2006,Croton2006,Fabian2012} or morphological quenching \citep{Martig2009}. Since our galaxies are massive, we believe that stellar feedback \citep{Hopkins2014} does not play a major role in their quenching.

In support of morphological quenching, we also find that our sample has a high $B/T$ ratio and \sersic index (e.g., median $n$ increasing from $3.49_{-0.15}^{+0.35}$ in F200W to $4.25_{-0.47}^{+1.18}$ in F444W, and median $B/T$ increasing from $0.66_{-0.07}^{+0.06}$ in F200W to $0.76_{-0.06}^{+0.02}$ in F444W), which has been linked to low sSFR \citep{Whitaker2015,Pan2018} and old stellar populations \citep{GonzalezDelgado2015,McDermid2015,LopezFernandez2018}. These types of galaxies tend to show inside-out quenching \citep{Lin2019}.

On the other hand, different models \citep{Hopkins2008,Somerville2008,Weinberger2017} and simulations \citep{Dubois2016,Weinberger2018,Dave2019} have shown that AGN feedback is necessary to quench the galaxies and to reproduce their observed properties. Inferences from simulations have also shown that AGN feedback in the form of kinetic winds is responsible for the suppression of central star formation and is required for inside-out quenching \citep{Nelson2019, Nelson2021}. 

\added[Ref. comment: soften AGN discussion]{While some galaxies in our sample exhibit faint diffraction-like patterns or central residuals in their structural fits, these features are very weak and cannot be robustly distinguished from systematic PSF modeling or deblending artifacts. Consequently, we find no direct, conclusive evidence of AGNs from our imaging data alone. To quantify the strength of any potential AGN contribution, we compute the \pysersic residuals for the central region of $0\farcs2\times0\farcs2$. In the F200W filter, the average absolute residual percentage is $6.73\% \pm 2.99\%$ with a maximum difference of $45.74\%$ (for \texttt{ID\_DR3} = 18351, which is difficult to model due to its disturbed morphology) and the next highest flux difference of $13.10\%$. In the F444W filter, the average absolute residual percentage is $1.45\% \pm 0.42\%$ with a maximum difference of $6.09\%$ and the next highest flux difference of $2.45\%$. This possible weak AGN contribution is not significant enough to affect our physical parameter derivations, consistent with the findings of \citet{Haryana_2025} that the impact of such weak AGNs on resolved SED fitting results is within $1\sigma$ uncertainties. For this reason, we do not include an AGN component in our \pixedfit SED modeling. Nonetheless, the observed inside-out quenching profiles (positive radial sSFR gradients) are similar to the profiles one expects from quenching driven by AGN feedback (e.g., via kinetic energy injection), a scenario widely predicted by cosmological simulations \citep{Nelson2019, Nelson2021}.}

\added{This theoretical picture aligns with recent spectroscopic observations that directly detect AGN-driven outflows and rapid quenching at cosmic noon. For instance, \citet{Park2024} found that massive quiescent galaxies at $z \sim 2$ show evidence of recent rapid quenching accompanied by central starbursts where AGN activity drives multi-phase gas outflows, while \citet{Belli2024} detected powerful neutral gas outflows at $z = 2.445$ with mass outflow rates sufficient to quench star formation. Furthermore, \citet{Scholtz2026} showed that central quenching can be a lengthy process driven by a meticulous balance between gas inflows and outflows regulated by AGN feedback. A more detailed analysis with spectroscopic follow-up using the JWST NIRSpec in its MSA or IFU mode will be required to search for direct evidence of outflows and to confirm whether any weak or obscured AGN are present in our sample.}

\begin{deluxetable}{lcccccc}
\tablecaption{Median S\'ersic Index, Axis Ratio, and Bulge-to-Total Ratio Across Filters\label{tab:median_n_q}}
\tablehead{
\colhead{Filter} & \colhead{} & \colhead{$ n_{\mathrm{med}} $} & \colhead{} & \colhead{$ q_{\mathrm{med}} $} & \colhead{} & \colhead{$B/T_{\mathrm{med}}$}
}
\startdata
F115W & & $1.73_{-0.39}^{+0.29}$ & & $0.71_{-0.06}^{+0.08}$ & & $0.54_{-0.11}^{+0.07}$ \\
F150W & & $2.35_{-0.43}^{+0.45}$ & & $0.74_{-0.06}^{+0.03}$ & & $0.63_{-0.08}^{+0.05}$ \\
F200W & & $3.49_{-0.15}^{+0.35}$ & & $0.75_{-0.05}^{+0.03}$ & & $0.66_{-0.07}^{+0.06}$ \\
F277W & & $4.20_{-0.29}^{+0.56}$ & & $0.62_{-0.02}^{+0.06}$ & & $0.67_{-0.02}^{+0.08}$ \\
F356W & & $5.43_{-0.72}^{+0.58}$ & & $0.65_{-0.04}^{+0.06}$ & & $0.75_{-0.06}^{+0.04}$ \\
F444W & & $4.25_{-0.47}^{+1.18}$ & & $0.64_{-0.04}^{+0.02}$ & & $0.76_{-0.06}^{+0.02}$ \\
\enddata
\tablecomments{Median S\'ersic index ($ n_{\mathrm{med}} $), axis ratio ($ q_{\mathrm{med}} = 1 - \epsilon_{\mathrm{bulge}} $), and bulge-to-total ratio ($B/T_{\mathrm{med}}$) calculated from massive quiescent galaxies at $z \sim 2$--$3$ with valid pysersic fits. Sample size varies by filter: F115W (12 galaxies), F150W--F444W (14 galaxies). Uncertainties represent 68\% confidence intervals ($\pm$1$\sigma$) derived from bootstrap resampling with 1000 iterations. Values shown as $\mathrm{median}_{-\Delta_{\mathrm{lo}}}^{+\Delta_{\mathrm{hi}}}$.}
\end{deluxetable}

\section{Summary \& Conclusions} \label{sec:conclusions}

We have presented a detailed analysis of the morphological and resolved star formation properties of massive quiescent galaxies at cosmic noon ($2 < z < 3$), using deep JWST NIRCam imaging from the UNCOVER Treasury program \citep{Bezanson2024} and MegaScience medium-band survey \citep{Suess2024} of the Abell 2744 lensing cluster field. Our main findings are summarized below.

\begin{enumerate}

\item Using \bagpipes \citep{Carnall2018}
SED modeling on the combined HST + JWST broadband and medium-band photometry, we select 17 massive quiescent galaxies with stellar masses $M_* \gtrsim 10^{10}$~M$_\odot$ and low specific star formation rates ($\mathrm{sSFR} < 0.2/t_\mathrm{age}$) in the redshift range $2 < z < 3$. Three of these are multiply-lensed images of the same source, and two are multiply-lensed images of another source \citep{Siegel2025, Furtak2023}, giving us 14 unique galaxies for our analysis.

\item Using \texttt{statmorph} \citep{Rodriguez-Gomez2019} and \texttt{pysersic} \citep{Pasha2023}, we perform non-parametric (CAS, Gini--$M_{20}$) and parametric (S\'ersic profile and bulge-disk decomposition) morphological analysis across all NIRCam broadband filters. Most galaxies in our sample are classified as intermediate type or S0s, with significant bulge components. The median S\'ersic index is around 4, consistent with bulge-dominated systems. This value remains relatively constant across a wide redshift range ($z \sim 1.5$--$4$) when compared with literature values, indicating that massive quiescent galaxies had a significant bulge component already in place at high redshifts \citep{Kawinwanichakij2026}.

\item The persistent $n \sim 4$ across cosmic time, combined with the bulge-dominated nature of our sample, demonstrates that the morphology of massive galaxies, and in particular the bulge component, is tightly linked to their quiescence across a wide redshift range. This extends the local-universe finding that morphology strongly correlates with sSFR regardless of environment \citep{Bait2017} to high redshifts, suggesting that the physical processes connecting bulge growth and star formation suppression have been operating since at least $z \sim 4$.

\item From our spatially resolved SED modeling using \texttt{piXedfit} \citep{Abdurrouf2021}, 11 out of the 14 ($\sim 79\%$) galaxies show positive specific star formation rate (sSFR) gradients, with lower sSFR in the center and higher sSFR in the outskirts. This is direct evidence that these galaxies are quenching from the inside out. The mean sSFR increases by roughly two orders of magnitude from $R/R_e = 0.5$ to $R/R_e = 4.5$. The remaining 3 galaxies do not show a clearly positive or negative sSFR gradient, instead showing off-center star-forming clumps or minor mergers. Two out of the 14 galaxies have distinct secondary star-forming cores.

\item The mean radial profiles of stellar mass, SFR, and sSFR are consistent with those reported by similar studies at comparable redshifts \citep{Haryana_2025, Laishram2025}. The formation time ($t_{50}$) radial profiles show that the inner regions ($< 4$ kpc) of the galaxies formed earlier than the outer regions ($> 4$ kpc) by approximately 0.5 Gyr, further supporting the inside-out formation and quenching scenario. The quenching timescale radial profiles indicate that the cores were quenched more rapidly than the outskirts. From \bagpipes SED modeling, the mean quenching timescale ($\Delta t_{BG} = t_{q,\,\mathrm{BG}} - t_{50,\,\mathrm{BG}}$) is $\approx 1.4$ Gyr, indicating rapid early mass assembly followed by efficient quenching. The mean half-mass radius of our sample is $R_e = 1.95 \pm 0.13$ kpc.

\item \added[Ref. comment: remove AGN claim from conclusion]{We find that the faint central residuals or diffraction spikes observed in some of our galaxies are statistically negligible (with average absolute residuals of $<2\%$ at F444W). The impact of any possible weak AGN contribution is within $1\sigma$ uncertainties of our resolved SED fitting results, confirming that our stellar population parameters are robust against point-source contamination. While these imaging features do not constitute direct evidence of AGNs, spectroscopic observations are required to confirm low-level central activity.}

\item \added[Ref. comment: soften AGN quenching driver in conclusions]{The observed inside-out quenching patterns and bulge-dominated morphologies collectively point toward morphological quenching \citep{Martig2009} and/or AGN feedback \citep{Bower2006, Croton2006, Fabian2012} as the primary drivers of quenching in these massive systems. The high \sersic indices and bulge-to-total ratios are linked to low sSFR and old stellar populations, consistent with the morphological quenching framework. Our resolved radial profiles test and are consistent with cosmological simulations that predict AGN feedback in the form of kinetic winds as a possible mechanism for suppressing central star formation and producing positive sSFR gradients \citep{Nelson2019, Nelson2021}.}

\end{enumerate}

Our results highlight the value of spatially resolved studies for understanding how quenching operates within galaxies at cosmic noon. The massive quiescent galaxies in our sample, with their established bulge components and inside-out quenching signatures, are similar to the massive elliptical galaxies we observe in the local universe. Future spectroscopic follow-up with JWST NIRSpec will be essential for confirming the role of AGN feedback through direct detection of outflows and for constraining the timescales of the quenching process in these early massive systems.  


\begin{acknowledgments}
\added{We thank the anonymous referee for their insightful comments that have helped improve both the content and presentation of this paper.} We are grateful to the UNCOVER and MegaScience teams for designing their observing programs, developing the data reduction pipelines, and providing the reduced data products to the public. 
This work is based on observations made with the NASA/ESA/CSA James Webb Space Telescope and the NASA/ESA Hubble Space Telescope. The data were obtained from the Mikulski Archive for Space Telescopes at the Space Telescope Science Institute, which is operated by the Association of Universities for Research in Astronomy, Inc., under NASA contract NAS 5-03127 for JWST and NAS 5-26555 for HST. These observations are associated with JWST Cycle 1 GO program 2561 and Cycle 2 GO program 4111, and gratefully made use of additional public JWST programs in the Abell 2744 field, including JWST-ERS-1324, JWST-DD-2756, JWST-GO-2641, JWST-GO-2883, JWST-GO-3516, and JWST-GO-3538. The relevant HST observations are associated with programs HST-GO-11689, HST-GO-13386, HST-GO-13389, HST-GO/DD-13495, HST-GO-15117, and HST-GO/DD-17231.
This work also made use of the DAWN JWST Archive (DJA). DJA is an initiative of the Cosmic Dawn Center, which is funded by the Danish National Research Foundation under grant DNRF140.
VP acknowledges support from the Indian Institute of Science Education and Research (IISER) Pune and the National Centre for Radio Astrophysics (NCRA-TIFR), Pune. RJ, YW, and PB acknowledge the support of the Department of Atomic Energy, Government of India, under the project title ``Next Generation Instrumentation for Radio Astronomy'', UID number RTI4017.
\end{acknowledgments}





\section{Data Availability}
\added{The results of our analysis are available on Zenodo with doi: \href{https://doi.org/10.5281/zenodo.21265814}{10.5281/zenodo.21265814}. It includes the outputs and plots of \texttt{BAGPIPES}, \texttt{statmorph}, \texttt{piXedfit}, and \texttt{pysersic} for all the 14 galaxies in our sample.} 

%
\facilities{HST, JWST}

\software{astropy \citep{2013A&A...558A..33A,2018AJ....156..123A,2022ApJ...935..167A},
          BAGPIPES \citep{Carnall2018},
          Gnuastro v0.23 \citep{gnuastro,astscript-color-faint-gray},
          piXedfit \citep{Abdurrouf2021},
          pysersic \citep{Pasha2023},
          scipy \citep{SciPy2020},
          sedpy \citep{johnson_2021_4582723},
          SEP \citep{Bertin1996, Barbary2016},
          statmorph \citep{Rodriguez-Gomez2019}
          }


\appendix
\section{Statmorph and pysersic results for the 14 galaxies in our sample} \label{sec:morph_results}

This section provides the comprehensive morphological analysis results for all 14 objects in our target sample. Table~\ref{tab:statmorph} summarizes the non-parametric morphological parameters, including concentration ($C$), asymmetry ($A$), smoothness ($S$), Gini coefficient, and $M_{20}$, derived using \statmorph. Table~\ref{tab:pysersic} presents the corresponding parametric \sersic fitting results obtained with \pysersic across multiple filters.

\startlongtable
\begin{deluxetable*}{lccccccc}
\tabletypesize{\scriptsize}
\tablecaption{Morphological parameters from statmorph analysis across multiple filters\label{tab:statmorph}}
\tablehead{
\colhead{Filter} & \colhead{ID} & \colhead{$C$} & \colhead{$A$} & \colhead{$S$} & \colhead{Gini} & \colhead{$M_{20}$} 
}
\startdata
F090W & 14207 & \nodata & \nodata & \nodata & \nodata & \nodata  \\
F115W &  & 3.76 & 0.196 & 0.000 & 0.700 & -2.43  \\
F150W &  & 3.50 & 0.141 & 0.072 & 0.593 & -1.88  \\
F200W &  & 3.34 & 0.093 & 0.072 & 0.574 & -1.93  \\
F277W &  & 2.99 & 0.058 & 0.000 & 0.537 & -1.76  \\
F356W &  & 3.01 & 0.059 & 0.000 & 0.530 & -1.82  \\
F444W &  & 3.00 & 0.054 & 0.095 & 0.538 & -1.79  \\
\hline
F090W & 14897 & 4.15 & 0.218 & 0.018 & 0.557 & -1.26  \\
F115W &  & 4.14 & 0.114 & -0.026 & 0.617 & -1.73  \\
F150W &  & 3.49 & 0.062 & 0.001 & 0.550 & -2.04  \\
F200W &  & 3.43 & 0.055 & 0.002 & 0.554 & -2.03  \\
F277W &  & 3.00 & 0.065 & 0.094 & 0.524 & -1.88  \\
F356W &  & 2.90 & 0.054 & 0.090 & 0.513 & -1.76  \\
F444W &  & 2.87 & 0.052 & 0.086 & 0.515 & -1.73  \\
\hline
F090W & 18351 & 3.80 & 0.052 & -2.588 & 0.642 & -2.04  \\
F115W &  & 4.35 & 0.160 & -1.085 & 0.691 & -1.77  \\
F150W &  & 4.76 & 0.267 & -0.163 & 0.674 & -1.80  \\
F200W &  & 4.71 & 0.227 & -0.023 & 0.679 & -2.02  \\
F277W &  & 4.22 & 0.155 & 0.028 & 0.657 & -2.25  \\
F356W &  & 4.18 & 0.185 & 0.037 & 0.635 & -2.17  \\
F444W &  & 4.13 & 0.241 & 0.041 & 0.612 & -1.80  \\
\hline
F090W & 20697 & \nodata & \nodata & \nodata & \nodata & \nodata  \\
F115W &  & 3.67 & -0.222 & -0.093 & 0.567 & -1.29  \\
F150W &  & 3.97 & 0.069 & -0.017 & 0.596 & -2.16  \\
F200W &  & 3.83 & 0.066 & -0.002 & 0.604 & -2.21  \\
F277W &  & 3.74 & 0.072 & 0.031 & 0.594 & -2.19  \\
F356W &  & 3.55 & 0.051 & 0.038 & 0.572 & -2.07  \\
F444W &  & 3.50 & 0.046 & 0.035 & 0.568 & -2.05  \\
\hline
F090W & 27482 & 3.12 & -0.125 & -0.045 & 0.549 & -1.91  \\
F115W &  & 3.40 & -0.003 & 0.010 & 0.589 & -2.05  \\
F150W &  & 3.28 & 0.063 & -0.001 & 0.542 & -1.94  \\
F200W &  & 3.20 & 0.040 & 0.001 & 0.544 & -1.92  \\
F277W &  & 2.82 & 0.048 & 0.000 & 0.510 & -1.78  \\
F356W &  & 2.80 & 0.040 & 0.000 & 0.507 & -1.80  \\
F444W &  & 2.81 & 0.039 & 0.096 & 0.522 & -1.74  \\
\hline
F090W & 29599 & 3.60 & -0.074 & -0.009 & 0.510 & -2.14  \\
F115W &  & 3.88 & -0.109 & -0.041 & 0.582 & -2.33  \\
F150W &  & 4.20 & 0.032 & -0.025 & 0.604 & -2.39  \\
F200W &  & 4.20 & 0.038 & -0.006 & 0.606 & -2.38  \\
F277W &  & 3.96 & 0.062 & 0.021 & 0.594 & -2.23  \\
F356W &  & 3.85 & 0.054 & 0.029 & 0.595 & -2.20  \\
F444W &  & 3.72 & 0.039 & 0.023 & 0.591 & -2.11  \\
\hline
F090W & 32926 & \nodata & \nodata & \nodata & \nodata & \nodata  \\
F115W &  & \nodata & \nodata & \nodata & \nodata & \nodata  \\
F150W &  & 4.19 & 0.067 & 0.001 & 0.631 & -2.24  \\
F200W &  & 3.82 & 0.059 & 0.025 & 0.584 & -2.10  \\
F277W &  & 3.50 & 0.068 & 0.061 & 0.589 & -1.92  \\
F356W &  & 3.40 & 0.059 & 0.058 & 0.583 & -1.90  \\
F444W &  & 3.18 & 0.052 & 0.064 & 0.542 & -1.84  \\
\hline
F090W & 34122 & 1.88 & -0.286 & -0.415 & 0.485 & -0.78  \\
F115W &  & 3.59 & -0.082 & -0.052 & 0.596 & -2.14  \\
F150W &  & 3.66 & 0.010 & -0.007 & 0.565 & -2.16  \\
F200W &  & 3.58 & 0.014 & -0.004 & 0.554 & -2.10  \\
F277W &  & 3.29 & 0.042 & 0.067 & 0.555 & -1.93  \\
F356W &  & 3.20 & 0.039 & 0.068 & 0.554 & -1.91  \\
F444W &  & 3.06 & 0.042 & 0.070 & 0.524 & -1.86  \\
\hline
F090W & 43489 & \nodata & \nodata & \nodata & \nodata & \nodata  \\
F115W &  & 2.56 & -0.170 & -0.086 & 0.482 & -1.30  \\
F150W &  & 2.94 & 0.038 & -0.008 & 0.538 & -1.80  \\
F200W &  & 3.18 & 0.071 & 0.000 & 0.533 & -2.00  \\
F277W &  & 3.57 & 0.056 & -0.001 & 0.568 & -2.12  \\
F356W &  & 3.65 & 0.045 & -0.001 & 0.577 & -2.15  \\
F444W &  & 3.61 & 0.033 & 0.029 & 0.574 & -2.10  \\
\hline
F090W & 45378 & 1.48 & -0.110 & -2.793 & 0.625 & -0.52  \\
F115W &  & 1.53 & 0.283 & 0.097 & 0.538 & -1.12  \\
F150W &  & 3.27 & 0.265 & 0.014 & 0.600 & -1.90  \\
F200W &  & 3.41 & 0.195 & 0.012 & 0.637 & -2.05  \\
F277W &  & 3.34 & 0.097 & 0.004 & 0.565 & -1.94  \\
F356W &  & 3.26 & 0.075 & 0.004 & 0.559 & -1.92  \\
F444W &  & 3.14 & 0.062 & 0.077 & 0.541 & -1.92  \\
\hline
F090W & 45398 & \nodata & \nodata & \nodata & \nodata & \nodata  \\
F115W &  & \nodata & \nodata & \nodata & \nodata & \nodata  \\
F150W &  & 3.72 & 0.046 & 0.021 & 0.574 & -2.15  \\
F200W &  & 3.64 & 0.035 & 0.027 & 0.578 & -2.10  \\
F277W &  & 3.25 & 0.086 & 0.074 & 0.561 & -1.89  \\
F356W &  & 3.13 & 0.063 & 0.076 & 0.546 & -1.80  \\
F444W &  & 3.06 & 0.052 & 0.077 & 0.531 & -1.81  \\
\hline
F090W & 48116 & \nodata & \nodata & \nodata & \nodata & \nodata  \\
F115W &  & 4.06 & 0.090 & 0.048 & 0.620 & -2.18  \\
F150W &  & 3.86 & 0.068 & 0.033 & 0.584 & -2.14  \\
F200W &  & 3.63 & 0.047 & 0.037 & 0.569 & -2.10  \\
F277W &  & 3.22 & 0.037 & 0.091 & 0.550 & -1.84  \\
F356W &  & 3.04 & 0.033 & 0.092 & 0.530 & -1.82  \\
F444W &  & 2.98 & 0.018 & 0.089 & 0.522 & -1.83  \\
\hline
F090W & 62921 & \nodata & \nodata & \nodata & \nodata & \nodata  \\
F115W &  & 3.36 & 0.075 & 0.061 & 0.555 & -1.97  \\
F150W &  & 3.36 & 0.074 & 0.067 & 0.572 & -1.99  \\
F200W &  & 3.25 & 0.069 & 0.003 & 0.563 & -1.95  \\
F277W &  & 2.84 & 0.089 & 0.000 & 0.516 & -1.74  \\
F356W &  & 2.85 & 0.086 & 0.000 & 0.520 & -1.82  \\
F444W &  & 2.89 & 0.075 & 0.095 & 0.535 & -1.75  \\
\hline
F090W & 68318 & \nodata & \nodata & \nodata & \nodata & \nodata  \\
F115W &  & 3.72 & 0.170 & 0.129 & 0.547 & -2.13  \\
F150W &  & 4.00 & 0.052 & 0.033 & 0.615 & -2.23  \\
F200W &  & 3.96 & 0.052 & 0.015 & 0.613 & -2.23  \\
F277W &  & 3.54 & 0.081 & 0.059 & 0.576 & -2.02  \\
F356W &  & 3.52 & 0.068 & 0.006 & 0.587 & -1.98  \\
F444W &  & 3.53 & 0.073 & 0.011 & 0.575 & -1.93  \\
\enddata
\tablecomments{$C$: concentration, $A$: asymmetry, $S$: clumpiness, $M_{20}$: second-order moment, $n$: S\'ersic index. \nodata\ indicates measurements unavailable or flagged as unreliable.}
\end{deluxetable*}

\startlongtable
\begin{deluxetable*}{rccccccc}
\tablecaption{Pysersic fit results across multiple filters\label{tab:pysersic}}
\tablehead{
\colhead{ID} & \colhead{Filter} & \colhead{$n$} & \colhead{$\theta$} & \colhead{$B/T$} & \colhead{$r_{\mathrm{eff},1}$} & \colhead{$r_{\mathrm{eff},2}$} & \colhead{$q(b/a)$} \\
\colhead{} & \colhead{} & \colhead{} & \colhead{(rad)} & \colhead{} & \colhead{(kpc)} & \colhead{(kpc)} & \colhead{}
}
\startdata
14207 & F090W & \nodata & \nodata & \nodata & \nodata & \nodata & \nodata \\
 & F115W & $1.96_{-0.20}^{+0.21}$ & $1.26_{-0.02}^{+0.02}$ & $0.51_{-0.02}^{+0.02}$ & $2.76_{-0.19}^{+0.19}$ & $0.47_{-0.01}^{+0.01}$ & $0.49_{-0.02}^{+0.02}$ \\
 & F150W & $4.85_{-0.12}^{+0.12}$ & $1.29_{-0.00}^{+0.00}$ & $0.68_{-0.01}^{+0.01}$ & $2.75_{-0.06}^{+0.08}$ & $0.79_{-0.01}^{+0.01}$ & $0.62_{-0.01}^{+0.01}$ \\
 & F200W & $5.61_{-0.07}^{+0.07}$ & $1.26_{-0.01}^{+0.01}$ & $0.68_{-0.00}^{+0.00}$ & $2.10_{-0.02}^{+0.02}$ & $0.97_{-0.01}^{+0.00}$ & $0.72_{-0.00}^{+0.00}$ \\
 & F277W & $7.96_{-0.05}^{+0.02}$ & $1.24_{-0.00}^{+0.00}$ & $0.80_{-0.01}^{+0.00}$ & $1.14_{-0.02}^{+0.02}$ & $1.08_{-0.01}^{+0.01}$ & $0.64_{-0.01}^{+0.01}$ \\
 & F356W & $7.69_{-0.16}^{+0.10}$ & $1.21_{-0.00}^{+0.00}$ & $0.69_{-0.01}^{+0.01}$ & $1.85_{-0.05}^{+0.05}$ & $1.10_{-0.01}^{+0.01}$ & $0.74_{-0.01}^{+0.01}$ \\
 & F444W & $7.95_{-0.06}^{+0.03}$ & $1.20_{-0.01}^{+0.01}$ & $0.77_{-0.00}^{+0.00}$ & $0.99_{-0.01}^{+0.01}$ & $1.08_{-0.02}^{+0.02}$ & $0.66_{-0.01}^{+0.01}$ \\
\hline
14897 & F090W & $1.54_{-0.10}^{+0.13}$ & $1.41_{-0.02}^{+0.02}$ & $0.33_{-0.02}^{+0.02}$ & $0.50_{-0.03}^{+0.04}$ & $3.83_{-0.21}^{+0.23}$ & $0.67_{-0.02}^{+0.02}$ \\
 & F115W & $1.50_{-0.05}^{+0.07}$ & $1.48_{-0.01}^{+0.01}$ & $0.43_{-0.01}^{+0.02}$ & $0.50_{-0.01}^{+0.02}$ & $2.52_{-0.07}^{+0.07}$ & $0.68_{-0.01}^{+0.01}$ \\
 & F150W & $2.60_{-0.05}^{+0.06}$ & $1.50_{-0.01}^{+0.01}$ & $0.68_{-0.01}^{+0.01}$ & $0.86_{-0.01}^{+0.01}$ & $1.97_{-0.03}^{+0.03}$ & $0.68_{-0.01}^{+0.01}$ \\
 & F200W & $3.54_{-0.05}^{+0.05}$ & $1.51_{-0.00}^{+0.00}$ & $0.64_{-0.00}^{+0.01}$ & $1.03_{-0.01}^{+0.01}$ & $1.53_{-0.01}^{+0.01}$ & $0.66_{-0.01}^{+0.00}$ \\
 & F277W & $3.55_{-0.07}^{+0.07}$ & $1.50_{-0.00}^{+0.00}$ & $0.52_{-0.01}^{+0.01}$ & $0.52_{-0.01}^{+0.01}$ & $1.65_{-0.01}^{+0.01}$ & $0.52_{-0.01}^{+0.01}$ \\
 & F356W & $3.62_{-0.08}^{+0.09}$ & $1.49_{-0.00}^{+0.00}$ & $0.51_{-0.01}^{+0.01}$ & $0.49_{-0.01}^{+0.01}$ & $1.61_{-0.01}^{+0.01}$ & $0.55_{-0.01}^{+0.01}$ \\
 & F444W & $3.78_{-0.14}^{+0.15}$ & $1.49_{-0.00}^{+0.00}$ & $0.49_{-0.01}^{+0.01}$ & $0.42_{-0.01}^{+0.01}$ & $1.57_{-0.01}^{+0.01}$ & $0.31_{-0.01}^{+0.01}$ \\
\hline
18351 & F090W & \nodata & \nodata & \nodata & \nodata & \nodata & \nodata \\
 & F115W & $0.82_{-0.02}^{+0.03}$ & $1.91_{-0.01}^{+0.01}$ & $0.22_{-0.00}^{+0.00}$ & $0.67_{-0.01}^{+0.01}$ & $7.28_{-0.17}^{+0.16}$ & $0.45_{-0.01}^{+0.01}$ \\
 & F150W & $3.00_{-0.10}^{+0.10}$ & $1.95_{-0.01}^{+0.01}$ & $0.92_{-0.01}^{+0.01}$ & $7.48_{-0.13}^{+0.15}$ & $4.41_{-0.28}^{+0.27}$ & $0.51_{-0.01}^{+0.01}$ \\
 & F200W & $3.53_{-0.18}^{+0.57}$ & $1.92_{-0.01}^{+0.01}$ & $0.95_{-0.00}^{+0.00}$ & $6.64_{-0.25}^{+0.24}$ & $0.47_{-0.13}^{+5.07}$ & $0.41_{-0.01}^{+0.05}$ \\
 & F277W & $7.25_{-0.23}^{+0.17}$ & $1.91_{-0.01}^{+0.01}$ & $0.65_{-0.00}^{+0.00}$ & $0.87_{-0.02}^{+0.02}$ & $8.19_{-0.13}^{+0.10}$ & $0.27_{-0.01}^{+0.01}$ \\
 & F356W & $7.97_{-0.04}^{+0.01}$ & $1.92_{-0.01}^{+0.01}$ & $0.73_{-0.00}^{+0.00}$ & $1.07_{-0.01}^{+0.01}$ & $8.38_{-0.13}^{+0.13}$ & $0.27_{-0.00}^{+0.00}$ \\
 & F444W & $7.98_{-0.03}^{+0.01}$ & $1.93_{-0.00}^{+0.00}$ & $0.76_{-0.00}^{+0.00}$ & $0.97_{-0.01}^{+0.01}$ & $8.32_{-0.10}^{+0.09}$ & $0.33_{-0.01}^{+0.01}$ \\
\hline
20697 & F090W & \nodata & \nodata & \nodata & \nodata & \nodata & \nodata \\
 & F115W & $4.02_{-0.25}^{+0.24}$ & $0.59_{-0.04}^{+0.04}$ & $0.59_{-0.05}^{+0.04}$ & $2.83_{-0.32}^{+0.29}$ & $5.43_{-0.27}^{+0.28}$ & $0.85_{-0.02}^{+0.03}$ \\
 & F150W & $5.15_{-0.08}^{+0.08}$ & $0.55_{-0.01}^{+0.01}$ & $0.74_{-0.01}^{+0.01}$ & $4.13_{-0.09}^{+0.10}$ & $4.29_{-0.08}^{+0.08}$ & $0.84_{-0.01}^{+0.01}$ \\
 & F200W & $4.62_{-0.07}^{+0.08}$ & $0.51_{-0.00}^{+0.00}$ & $0.82_{-0.01}^{+0.01}$ & $3.49_{-0.05}^{+0.05}$ & $3.95_{-0.06}^{+0.06}$ & $0.78_{-0.00}^{+0.00}$ \\
 & F277W & $4.30_{-0.03}^{+0.04}$ & $0.39_{-0.00}^{+0.00}$ & $0.87_{-0.00}^{+0.00}$ & $3.96_{-0.03}^{+0.03}$ & $1.69_{-0.02}^{+0.03}$ & $0.62_{-0.00}^{+0.00}$ \\
 & F356W & $5.76_{-0.04}^{+0.04}$ & $0.37_{-0.00}^{+0.00}$ & $0.87_{-0.00}^{+0.00}$ & $3.45_{-0.03}^{+0.02}$ & $2.05_{-0.02}^{+0.02}$ & $0.60_{-0.00}^{+0.00}$ \\
 & F444W & $5.43_{-0.07}^{+0.07}$ & $0.37_{-0.00}^{+0.00}$ & $0.88_{-0.00}^{+0.00}$ & $3.28_{-0.03}^{+0.03}$ & $1.86_{-0.02}^{+0.02}$ & $0.59_{-0.00}^{+0.00}$ \\
\hline
27482 & F090W & \nodata & \nodata & \nodata & \nodata & \nodata & \nodata \\
 & F115W & $1.99_{-0.15}^{+0.15}$ & $3.12_{-0.02}^{+0.01}$ & $0.86_{-0.03}^{+0.03}$ & $1.28_{-0.07}^{+0.07}$ & $0.48_{-0.04}^{+0.04}$ & $0.79_{-0.02}^{+0.02}$ \\
 & F150W & $2.80_{-0.07}^{+0.07}$ & $0.02_{-0.01}^{+0.01}$ & $0.66_{-0.01}^{+0.01}$ & $1.20_{-0.01}^{+0.01}$ & $0.83_{-0.01}^{+0.01}$ & $0.82_{-0.01}^{+0.01}$ \\
 & F200W & $3.39_{-0.06}^{+0.06}$ & $0.07_{-0.01}^{+0.01}$ & $0.59_{-0.01}^{+0.01}$ & $1.14_{-0.01}^{+0.01}$ & $0.84_{-0.00}^{+0.00}$ & $0.82_{-0.01}^{+0.01}$ \\
 & F277W & $4.10_{-0.12}^{+0.13}$ & $0.09_{-0.01}^{+0.01}$ & $0.58_{-0.01}^{+0.01}$ & $0.47_{-0.01}^{+0.01}$ & $0.95_{-0.01}^{+0.01}$ & $0.73_{-0.01}^{+0.01}$ \\
 & F356W & $3.83_{-0.06}^{+0.07}$ & $0.17_{-0.00}^{+0.00}$ & $0.64_{-0.01}^{+0.01}$ & $0.52_{-0.00}^{+0.00}$ & $0.85_{-0.01}^{+0.01}$ & $0.67_{-0.01}^{+0.01}$ \\
 & F444W & $2.99_{-0.04}^{+0.05}$ & $0.15_{-0.01}^{+0.02}$ & $0.73_{-0.01}^{+0.01}$ & $0.50_{-0.00}^{+0.00}$ & $1.01_{-0.01}^{+0.01}$ & $0.79_{-0.01}^{+0.01}$ \\
\hline
29599 & F090W & \nodata & \nodata & \nodata & \nodata & \nodata & \nodata \\
 & F115W & \nodata & \nodata & \nodata & \nodata & \nodata & \nodata \\
 & F150W & $1.74_{-0.06}^{+0.06}$ & $1.90_{-0.01}^{+0.01}$ & $0.84_{-0.00}^{+0.00}$ & $3.00_{-0.04}^{+0.05}$ & $0.30_{-0.00}^{+0.01}$ & $0.67_{-0.01}^{+0.01}$ \\
 & F200W & $2.01_{-0.06}^{+0.06}$ & $1.93_{-0.01}^{+0.01}$ & $0.84_{-0.01}^{+0.01}$ & $2.93_{-0.04}^{+0.04}$ & $0.29_{-0.01}^{+0.01}$ & $0.68_{-0.00}^{+0.00}$ \\
 & F277W & $3.48_{-0.13}^{+0.15}$ & $1.81_{-0.01}^{+0.01}$ & $0.49_{-0.00}^{+0.00}$ & $0.50_{-0.01}^{+0.01}$ & $4.51_{-0.04}^{+0.05}$ & $0.83_{-0.01}^{+0.01}$ \\
 & F356W & $3.64_{-0.14}^{+0.15}$ & $1.79_{-0.01}^{+0.00}$ & $0.53_{-0.00}^{+0.01}$ & $0.48_{-0.01}^{+0.01}$ & $4.66_{-0.04}^{+0.05}$ & $0.87_{-0.01}^{+0.01}$ \\
 & F444W & $3.16_{-0.20}^{+0.21}$ & $1.82_{-0.01}^{+0.01}$ & $0.52_{-0.01}^{+0.01}$ & $0.45_{-0.01}^{+0.01}$ & $4.36_{-0.05}^{+0.05}$ & $0.90_{-0.02}^{+0.01}$ \\
\hline
32926 & F090W & \nodata & \nodata & \nodata & \nodata & \nodata & \nodata \\
 & F115W & \nodata & \nodata & \nodata & \nodata & \nodata & \nodata \\
 & F150W & $2.00_{-0.09}^{+0.11}$ & $2.58_{-0.06}^{+0.06}$ & $0.51_{-0.01}^{+0.01}$ & $0.43_{-0.01}^{+0.02}$ & $2.24_{-0.07}^{+0.08}$ & $0.88_{-0.01}^{+0.01}$ \\
 & F200W & $2.13_{-0.09}^{+0.12}$ & $2.76_{-0.08}^{+0.07}$ & $0.55_{-0.01}^{+0.01}$ & $0.49_{-0.01}^{+0.01}$ & $2.22_{-0.04}^{+0.04}$ & $0.94_{-0.01}^{+0.01}$ \\
 & F277W & $4.49_{-0.27}^{+0.28}$ & $2.34_{-1.05}^{+0.38}$ & $0.65_{-0.00}^{+0.00}$ & $0.38_{-0.01}^{+0.00}$ & $2.22_{-0.07}^{+0.06}$ & $0.98_{-0.01}^{+0.01}$ \\
 & F356W & $6.40_{-0.26}^{+0.26}$ & $2.07_{-0.97}^{+0.49}$ & $0.78_{-0.00}^{+0.00}$ & $0.54_{-0.01}^{+0.01}$ & $2.21_{-0.09}^{+0.08}$ & $0.99_{-0.01}^{+0.00}$ \\
 & F444W & $3.96_{-0.53}^{+0.53}$ & $0.37_{-0.25}^{+2.58}$ & $0.64_{-0.02}^{+0.02}$ & $0.32_{-0.02}^{+0.02}$ & $2.11_{-0.06}^{+0.06}$ & $0.95_{-0.02}^{+0.01}$ \\
\hline
34122 & F090W & \nodata & \nodata & \nodata & \nodata & \nodata & \nodata \\
 & F115W & $0.97_{-0.09}^{+0.10}$ & $0.19_{-0.03}^{+0.03}$ & $0.21_{-0.02}^{+0.02}$ & $0.42_{-0.03}^{+0.03}$ & $2.35_{-0.09}^{+0.08}$ & $0.73_{-0.03}^{+0.03}$ \\
 & F150W & $1.20_{-0.04}^{+0.04}$ & $0.22_{-0.01}^{+0.01}$ & $0.26_{-0.01}^{+0.01}$ & $0.45_{-0.01}^{+0.01}$ & $2.23_{-0.02}^{+0.02}$ & $0.77_{-0.01}^{+0.01}$ \\
 & F200W & $3.44_{-0.05}^{+0.05}$ & $0.19_{-0.01}^{+0.01}$ & $0.69_{-0.01}^{+0.01}$ & $1.16_{-0.01}^{+0.02}$ & $2.26_{-0.03}^{+0.02}$ & $0.78_{-0.01}^{+0.01}$ \\
 & F277W & $5.42_{-0.20}^{+0.20}$ & $0.21_{-0.01}^{+0.00}$ & $0.62_{-0.00}^{+0.00}$ & $0.73_{-0.01}^{+0.01}$ & $2.00_{-0.02}^{+0.02}$ & $0.76_{-0.01}^{+0.01}$ \\
 & F356W & $5.27_{-0.14}^{+0.14}$ & $0.21_{-0.00}^{+0.00}$ & $0.66_{-0.00}^{+0.00}$ & $0.72_{-0.01}^{+0.01}$ & $2.07_{-0.02}^{+0.02}$ & $0.87_{-0.01}^{+0.01}$ \\
 & F444W & $4.55_{-0.19}^{+0.18}$ & $0.21_{-0.01}^{+0.00}$ & $0.65_{-0.01}^{+0.01}$ & $0.66_{-0.02}^{+0.02}$ & $2.05_{-0.02}^{+0.02}$ & $0.84_{-0.02}^{+0.01}$ \\
\hline
43489 & F090W & \nodata & \nodata & \nodata & \nodata & \nodata & \nodata \\
 & F115W & $4.45_{-0.83}^{+0.88}$ & $0.19_{-0.03}^{+0.04}$ & $0.06_{-0.02}^{+0.03}$ & $8.35_{-1.04}^{+1.24}$ & $4.10_{-0.12}^{+0.11}$ & $0.80_{-0.19}^{+0.14}$ \\
 & F150W & $1.74_{-0.10}^{+0.11}$ & $0.22_{-0.01}^{+0.01}$ & $0.23_{-0.02}^{+0.02}$ & $2.45_{-0.16}^{+0.16}$ & $4.93_{-0.07}^{+0.07}$ & $0.99_{-0.01}^{+0.01}$ \\
 & F200W & $4.14_{-0.13}^{+0.14}$ & $0.17_{-0.00}^{+0.00}$ & $0.35_{-0.01}^{+0.01}$ & $3.54_{-0.12}^{+0.12}$ & $4.09_{-0.03}^{+0.03}$ & $0.92_{-0.02}^{+0.02}$ \\
 & F277W & $3.88_{-0.10}^{+0.10}$ & $0.16_{-0.00}^{+0.00}$ & $0.76_{-0.00}^{+0.00}$ & $5.60_{-0.13}^{+0.10}$ & $2.64_{-0.06}^{+0.07}$ & $0.60_{-0.00}^{+0.00}$ \\
 & F356W & $4.61_{-0.10}^{+0.10}$ & $0.15_{-0.00}^{+0.00}$ & $0.82_{-0.00}^{+0.00}$ & $4.55_{-0.07}^{+0.06}$ & $2.61_{-0.05}^{+0.05}$ & $0.62_{-0.00}^{+0.00}$ \\
 & F444W & $6.95_{-0.13}^{+0.11}$ & $0.15_{-0.00}^{+0.00}$ & $0.83_{-0.00}^{+0.00}$ & $4.55_{-0.05}^{+0.06}$ & $2.71_{-0.03}^{+0.03}$ & $0.62_{-0.00}^{+0.00}$ \\
\hline
45378 & F090W & \nodata & \nodata & \nodata & \nodata & \nodata & \nodata \\
 & F115W & $1.06_{-0.10}^{+0.10}$ & $0.60_{-0.01}^{+0.01}$ & $0.63_{-0.03}^{+0.02}$ & $5.54_{-0.28}^{+0.35}$ & $2.69_{-0.08}^{+0.08}$ & $0.61_{-0.03}^{+0.03}$ \\
 & F150W & $0.65_{-0.00}^{+0.00}$ & $0.57_{-0.00}^{+0.00}$ & $0.28_{-0.00}^{+0.00}$ & $1.90_{-0.02}^{+0.02}$ & $5.30_{-0.06}^{+0.05}$ & $0.31_{-0.00}^{+0.00}$ \\
 & F200W & $0.70_{-0.01}^{+0.01}$ & $0.53_{-0.00}^{+0.00}$ & $0.36_{-0.00}^{+0.00}$ & $1.64_{-0.02}^{+0.02}$ & $5.31_{-0.08}^{+0.06}$ & $0.41_{-0.00}^{+0.00}$ \\
 & F277W & $2.25_{-0.02}^{+0.02}$ & $0.53_{-0.00}^{+0.00}$ & $0.80_{-0.00}^{+0.00}$ & $2.97_{-0.01}^{+0.01}$ & $1.28_{-0.01}^{+0.01}$ & $0.48_{-0.00}^{+0.00}$ \\
 & F356W & $2.69_{-0.02}^{+0.03}$ & $0.51_{-0.00}^{+0.00}$ & $0.93_{-0.00}^{+0.00}$ & $1.80_{-0.00}^{+0.00}$ & $4.42_{-0.13}^{+0.12}$ & $0.43_{-0.00}^{+0.00}$ \\
 & F444W & $3.24_{-0.02}^{+0.02}$ & $0.49_{-0.00}^{+0.00}$ & $0.87_{-0.00}^{+0.00}$ & $1.82_{-0.00}^{+0.00}$ & $1.83_{-0.02}^{+0.02}$ & $0.47_{-0.00}^{+0.00}$ \\
\hline
45398 & F090W & \nodata & \nodata & \nodata & \nodata & \nodata & \nodata \\
 & F115W & $1.35_{-0.14}^{+0.18}$ & $0.39_{-0.05}^{+0.05}$ & $0.71_{-0.03}^{+0.02}$ & $2.06_{-0.12}^{+0.15}$ & $0.33_{-0.01}^{+0.01}$ & $0.68_{-0.03}^{+0.03}$ \\
 & F150W & $2.10_{-0.09}^{+0.09}$ & $0.36_{-0.01}^{+0.01}$ & $0.60_{-0.02}^{+0.02}$ & $0.62_{-0.02}^{+0.03}$ & $3.07_{-0.10}^{+0.10}$ & $0.85_{-0.01}^{+0.01}$ \\
 & F200W & $2.83_{-0.04}^{+0.04}$ & $0.36_{-0.01}^{+0.01}$ & $0.77_{-0.01}^{+0.01}$ & $0.89_{-0.01}^{+0.01}$ & $3.20_{-0.08}^{+0.08}$ & $0.82_{-0.00}^{+0.00}$ \\
 & F277W & $3.57_{-0.14}^{+0.13}$ & $0.39_{-0.01}^{+0.01}$ & $0.64_{-0.00}^{+0.00}$ & $0.49_{-0.01}^{+0.01}$ & $2.26_{-0.03}^{+0.03}$ & $0.77_{-0.01}^{+0.01}$ \\
 & F356W & $4.71_{-0.09}^{+0.08}$ & $0.38_{-0.00}^{+0.00}$ & $0.80_{-0.00}^{+0.00}$ & $0.77_{-0.01}^{+0.01}$ & $2.12_{-0.03}^{+0.03}$ & $0.73_{-0.01}^{+0.01}$ \\
 & F444W & $3.79_{-0.06}^{+0.06}$ & $0.37_{-0.01}^{+0.01}$ & $0.79_{-0.00}^{+0.00}$ & $0.73_{-0.01}^{+0.01}$ & $2.27_{-0.04}^{+0.04}$ & $0.66_{-0.00}^{+0.00}$ \\
\hline
48116 & F090W & $3.44_{-0.17}^{+0.18}$ & $0.17_{-0.02}^{+0.02}$ & $0.57_{-0.01}^{+0.01}$ & $0.35_{-0.01}^{+0.01}$ & $7.61_{-0.53}^{+0.53}$ & $0.96_{-0.02}^{+0.01}$ \\
 & F115W & $2.50_{-0.06}^{+0.06}$ & $0.44_{-0.02}^{+0.02}$ & $0.63_{-0.01}^{+0.00}$ & $0.47_{-0.01}^{+0.01}$ & $5.08_{-0.22}^{+0.22}$ & $0.86_{-0.01}^{+0.01}$ \\
 & F150W & $3.56_{-0.06}^{+0.06}$ & $0.58_{-0.01}^{+0.00}$ & $0.84_{-0.00}^{+0.00}$ & $2.01_{-0.02}^{+0.02}$ & $0.47_{-0.01}^{+0.01}$ & $0.65_{-0.00}^{+0.00}$ \\
 & F200W & $4.15_{-0.04}^{+0.04}$ & $0.56_{-0.00}^{+0.00}$ & $0.81_{-0.00}^{+0.00}$ & $1.93_{-0.01}^{+0.02}$ & $0.80_{-0.01}^{+0.01}$ & $0.62_{-0.00}^{+0.00}$ \\
 & F277W & $5.42_{-0.07}^{+0.07}$ & $0.56_{-0.00}^{+0.00}$ & $0.87_{-0.00}^{+0.00}$ & $1.12_{-0.00}^{+0.00}$ & $1.43_{-0.02}^{+0.02}$ & $0.59_{-0.00}^{+0.00}$ \\
 & F356W & $6.27_{-0.08}^{+0.08}$ & $0.55_{-0.00}^{+0.00}$ & $0.82_{-0.00}^{+0.00}$ & $1.05_{-0.01}^{+0.01}$ & $1.39_{-0.02}^{+0.02}$ & $0.62_{-0.00}^{+0.00}$ \\
 & F444W & $7.34_{-0.07}^{+0.07}$ & $0.57_{-0.00}^{+0.00}$ & $0.76_{-0.00}^{+0.00}$ & $0.95_{-0.00}^{+0.00}$ & $1.42_{-0.01}^{+0.01}$ & $0.62_{-0.00}^{+0.00}$ \\
\hline
62921 & F090W & $2.35_{-0.11}^{+0.11}$ & $1.42_{-0.02}^{+0.01}$ & $0.49_{-0.06}^{+0.05}$ & $0.95_{-0.03}^{+0.03}$ & $68.90_{-3.13}^{+3.38}$ & $0.56_{-0.01}^{+0.01}$ \\
 & F115W & $2.05_{-0.04}^{+0.04}$ & $1.44_{-0.01}^{+0.01}$ & $0.58_{-0.06}^{+0.04}$ & $0.88_{-0.01}^{+0.01}$ & $68.75_{-2.89}^{+3.02}$ & $0.54_{-0.00}^{+0.00}$ \\
 & F150W & $5.80_{-0.20}^{+0.23}$ & $1.44_{-0.01}^{+0.00}$ & $0.59_{-0.00}^{+0.00}$ & $1.61_{-0.07}^{+0.09}$ & $1.01_{-0.01}^{+0.01}$ & $0.72_{-0.01}^{+0.01}$ \\
 & F200W & $6.00_{-0.15}^{+0.17}$ & $1.45_{-0.00}^{+0.00}$ & $0.58_{-0.00}^{+0.00}$ & $1.54_{-0.03}^{+0.04}$ & $1.04_{-0.01}^{+0.01}$ & $0.76_{-0.01}^{+0.01}$ \\
 & F277W & $7.97_{-0.04}^{+0.02}$ & $1.48_{-0.00}^{+0.00}$ & $0.68_{-0.00}^{+0.00}$ & $0.71_{-0.01}^{+0.01}$ & $1.13_{-0.01}^{+0.01}$ & $0.63_{-0.01}^{+0.01}$ \\
 & F356W & $7.97_{-0.04}^{+0.02}$ & $1.46_{-0.00}^{+0.00}$ & $0.63_{-0.01}^{+0.00}$ & $0.76_{-0.01}^{+0.02}$ & $1.07_{-0.01}^{+0.01}$ & $0.67_{-0.01}^{+0.01}$ \\
 & F444W & $7.89_{-0.10}^{+0.05}$ & $1.48_{-0.00}^{+0.00}$ & $0.59_{-0.01}^{+0.01}$ & $0.77_{-0.02}^{+0.02}$ & $1.01_{-0.01}^{+0.01}$ & $0.71_{-0.01}^{+0.01}$ \\
\hline
68318 & F090W & \nodata & \nodata & \nodata & \nodata & \nodata & \nodata \\
 & F115W & $0.81_{-0.07}^{+0.11}$ & $1.64_{-0.05}^{+0.05}$ & $0.25_{-0.03}^{+0.03}$ & $0.44_{-0.04}^{+0.04}$ & $2.86_{-0.21}^{+0.21}$ & $0.82_{-0.05}^{+0.04}$ \\
 & F150W & $1.42_{-0.06}^{+0.07}$ & $1.62_{-0.02}^{+0.02}$ & $0.40_{-0.01}^{+0.01}$ & $0.58_{-0.02}^{+0.02}$ & $3.48_{-0.10}^{+0.10}$ & $0.76_{-0.01}^{+0.02}$ \\
 & F200W & $1.15_{-0.04}^{+0.04}$ & $1.65_{-0.01}^{+0.01}$ & $0.38_{-0.01}^{+0.01}$ & $0.50_{-0.01}^{+0.01}$ & $2.88_{-0.06}^{+0.06}$ & $0.75_{-0.01}^{+0.01}$ \\
 & F277W & $3.94_{-0.19}^{+0.19}$ & $1.60_{-0.01}^{+0.01}$ & $0.74_{-0.01}^{+0.01}$ & $1.15_{-0.03}^{+0.03}$ & $3.35_{-0.14}^{+0.18}$ & $0.61_{-0.01}^{+0.01}$ \\
 & F356W & $5.59_{-0.20}^{+0.18}$ & $1.58_{-0.01}^{+0.01}$ & $0.77_{-0.01}^{+0.01}$ & $0.85_{-0.02}^{+0.02}$ & $3.47_{-0.11}^{+0.14}$ & $0.60_{-0.01}^{+0.01}$ \\
 & F444W & $2.83_{-0.17}^{+0.16}$ & $1.59_{-0.01}^{+0.01}$ & $0.79_{-0.01}^{+0.01}$ & $2.16_{-0.06}^{+0.05}$ & $0.19_{-0.01}^{+0.02}$ & $0.58_{-0.01}^{+0.01}$ \\
\enddata
\tablecomments{Values shown as median with 16th/84th percentile uncertainties: $\mathrm{value}_{-\sigma_{\mathrm{lo}}}^{+\sigma_{\mathrm{hi}}}$. $n$: S\'ersic index, $\theta$: position angle (rad), $B/T$: bulge to total luminosity ratio, $r_{\mathrm{eff},1/2}$: effective radii (kpc), $q(b/a)$: axis ratio.}
\end{deluxetable*}

\section{The three lensed galaxies (Source 67)} \label{sec:lensed_galaxies}

The three lensed galaxies, \texttt{ID\_DR3} = 45356, 45357, and 45378, do not have reliable SED modeling results due to bimodal \bagpipes posteriors in some parameters. We include the analysis of \texttt{ID\_DR3} = 45378 since it has been confirmed to be a massive quiescent galaxy by \citet{Siegel2025}. 

\begin{figure*}[ht!]
\centering
\includegraphics[width=0.3\textwidth]{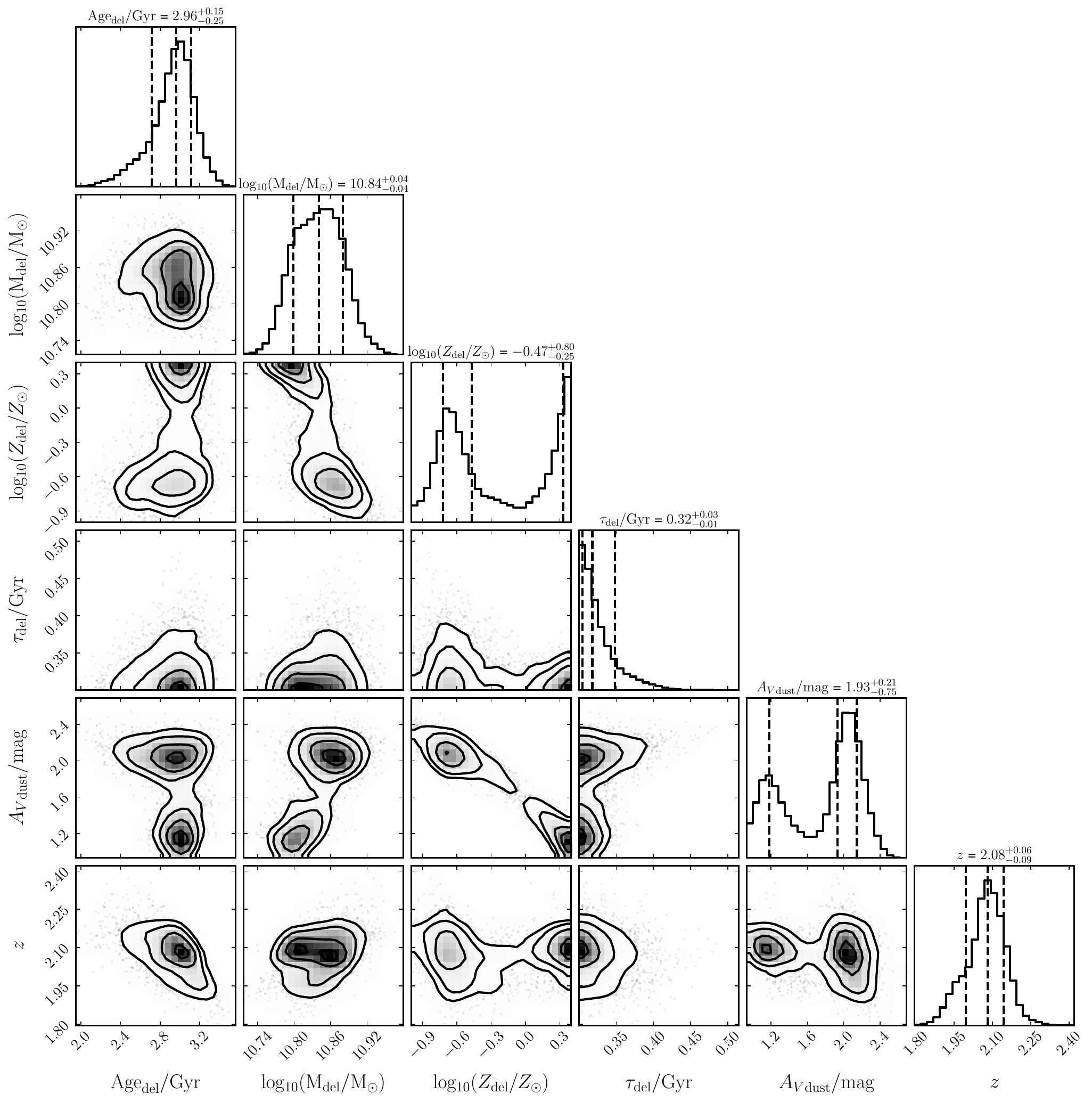}
\includegraphics[width=0.3\textwidth]{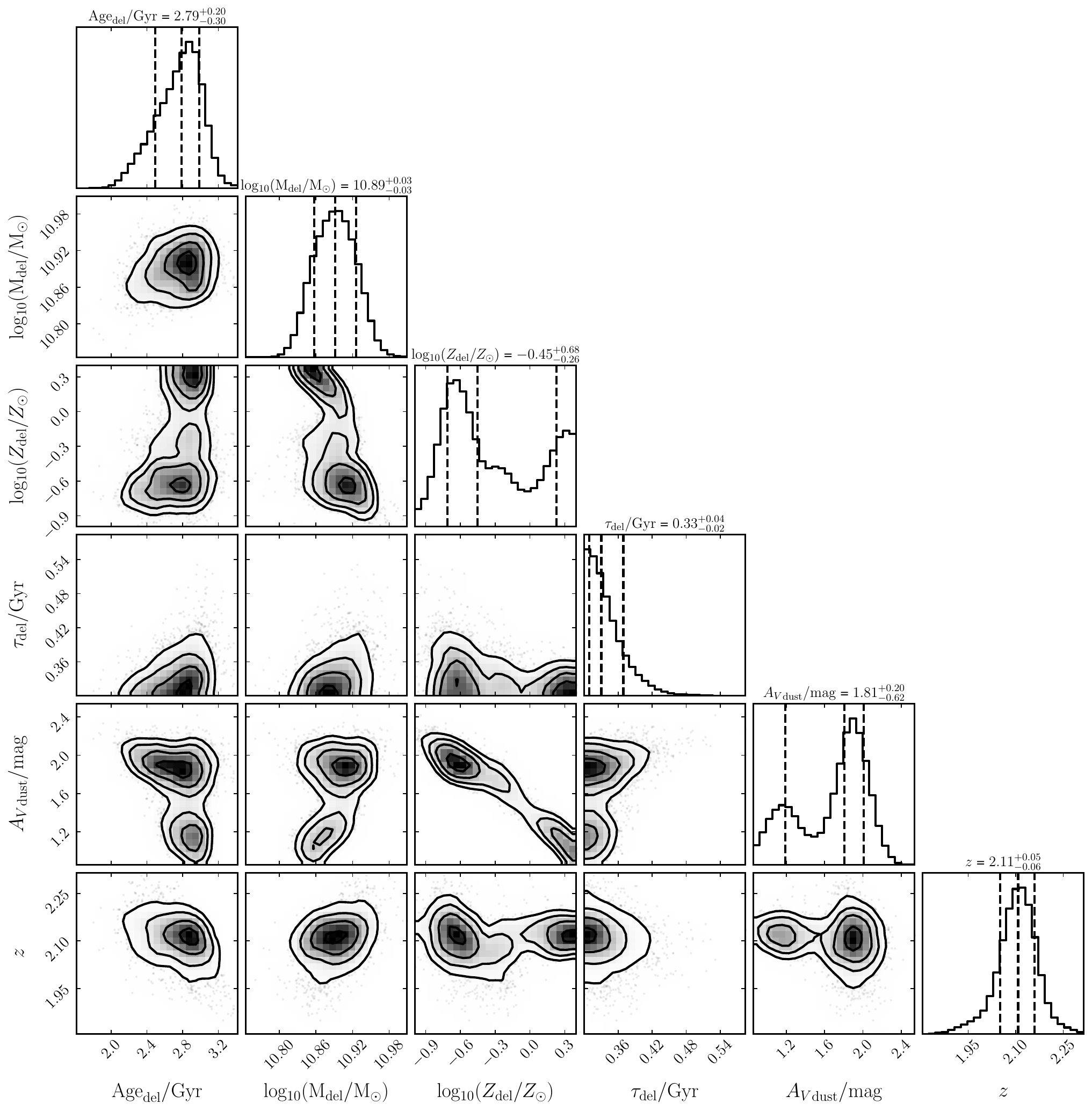}
\includegraphics[width=0.3\textwidth]{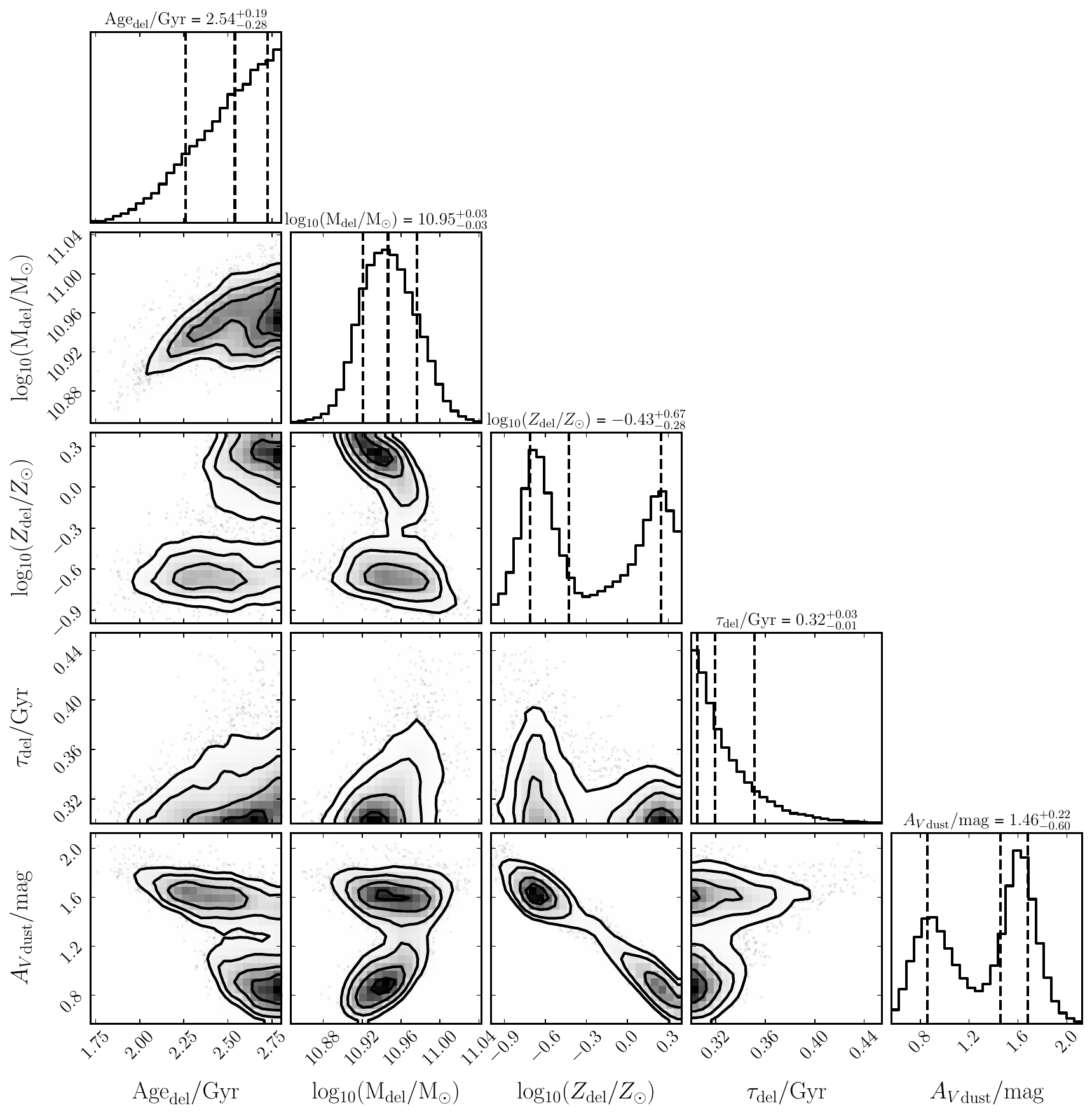}
\caption{\bagpipes posterior corner plots for the three lensed images of the same galaxy: \texttt{ID\_DR3} = 45356 (left), 45357 (center), and 45378 (right). The bimodal posteriors in metallicity and dust ($A_V$) are clearly visible.}\label{fig:lensed_corners}
\end{figure*}

Figure~\ref{fig:lensed_corners} shows the \bagpipes posterior corner plots for each of the three lensed images, illustrating the bimodal posteriors leading to unreliable parameter estimates.

\bibliographystyle{aasjournalv8}
\bibliography{references}{}



\end{document}